\documentclass[manuscript,screen]{acmart}

\usepackage[utf8]{inputenc}
\usepackage{booktabs} 
\usepackage{amssymb} 
\usepackage{graphicx}
\usepackage{subcaption}
\usepackage{rotating}
\usepackage{adjustbox}
\usepackage{stfloats}
\usepackage{array}
\usepackage[normalem]{ulem}
\usepackage{enumitem}
\usepackage{multirow}
\usepackage{tikz}
\usetikzlibrary{positioning,calc,arrows.meta}
\usepackage{multicol}
\usepackage[table,xcdraw]{xcolor}
\usepackage{caption,subcaption}
\usepackage{color,soul}
\usepackage{float}
\usepackage{xcolor} 
\usepackage{tabularx}
\usepackage{colortbl}
\usepackage{soul}
\usepackage{amsmath,bm}
\newcommand\crule[3][black]{\textcolor{#1}{\rule{#2}{#3}}}
\AtBeginDocument{%
  }

\setcopyright{acmlicensed}
\copyrightyear{2025}
\acmYear{2025}
\acmDOI{XXXXXXX.XXXXXXX}
\acmBooktitle{ACM Transactions on Recommender Systems (TORS)}
\acmISBN{978-1-4503-XXXX-X/18/06}

\begin{document}

\title{Investigating the Effects of Different Levels of User Control in an Interactive Educational Recommender System} 





\author{Qurat Ul Ain}
\email{qurat.ain@stud.uni-due.de}

\author{Mohamed Amine Chatti}
\email{mohamed.chatti@uni-due.de}

\author{William Kana Tsoplefack}
\email{william.kana-tsoplefack@stud.uni-due.de}

\author{Rawaa Alatrash}
\email{rawaa.alatrash@stud.uni-due.de}

\author{Shoeb Joarder}
\email{shoeb.joarder@uni-due.de}

\affiliation{
\institution{Social Computing Group, Faculty of Computer Science, University of Duisburg-Essen}
\city{Duisburg}
\country{Germany}
}

\renewcommand{\shortauthors}{Ain et al.}

\begin{abstract} 
Educational recommender systems (ERSs) are becoming increasingly important in enhancing educational outcomes and personalizing learning experiences by providing recommendations of personalized resources and activities to learners, tailored to their individual learning needs.  While user control is widely assumed to improve user experience, the effects of different levels of control in ERSs remain underexplored. To address this gap, we designed and evaluated an interactive ERS within the MOOC platform CourseMapper, where learners could interact with the input (i.e., user profile), process (i.e., recommendation algorithm), and output (i.e., recommendations) of the system. We conducted a between-subjects user study (N=184) to examine how varying levels of user control in an ERS influenced users’ perceptions of the recommendation goals of perceived control, transparency, trust, satisfaction, and perceived quality.
Our results show that enabling users to build and refine their profile is sufficient to promote positive perceptions of the ERS, while additional control options mainly reinforce these impressions. Moreover, perceived control is the only goal significantly affected by providing different levels of user control in the ERS, with input control exerting the strongest influence. Furthermore, different levels of control affect transparency, trust, satisfaction, and perceived quality in distinct yet interconnected ways.  Overall, the findings provide empirical evidence that user control positively shapes transparency, trust, satisfaction, and perceived quality, though to varying extents.
\end{abstract}




\maketitle

\section{Introduction}
Recommender systems (RS) are increasingly being utilized across
various domains. These systems have proven effective at enhancing user experience and aiding decision-making through personalized recommendations. In recent decades, RSs have also been applied to the field of education, leading to the development of educational recommender systems (ERS) \cite{manouselis2011recommender, khanal2020systematic}. In this context, ERSs are for example used to create personalized learning experiences \cite{valtolina2024design}, recommend suitable formal or informal learning materials \cite{chau2018learning}, suggest MOOCs \cite{BOUSBAHI20151813}, and adapt to context-aware learning environments \cite{santos2016toward}.

Traditional RSs typically offer limited interaction options in their user interfaces, allowing users only to express whether they like or dislike a recommendation \cite{jin2018effects}. Interactive RSs (IntRSs) have gained attention as an approach to empower users to control and interact with the RS \cite{he2016interactive, jugovac2017interacting, jannach2017user}. Controllability refers to the extent to which users can modify the RS to improve the quality of recommendations \cite{he2016interactive}. Concretely, the users can control the RS at three different levels, namely interacting with the input (i.e., user profile), process (i.e., recommendation algorithm), or output (i.e., recommendations) \cite{he2016interactive,harambam}. Therefore, controllability serves as an important indicator of the overall user experience in RSs, with lower levels of user control adversely affecting perceived recommendation quality \cite{jin2017different}. The control is provided in the RSs by allowing users to adjust preferences, change parameters of the underlying algorithm, directly interact with recommendations, and provide feedback, resulting in greater perceived control and a more transparent recommendation process \cite{knijnenburg2012inspectability, knijnenburg2012explaining}.
However, while IntRSs are effectively utilized in e-commerce, entertainment, and social media, the aspect of user control in educational recommender systems (ERSs) remains under-explored. 

User control has been demonstrated to positively affect various recommendation goals, including perceived accuracy of recommendations \cite{schaffer2015hypothetical, Donovan, 2012tasteweights, maxwell2015, smallworlds}, usability \cite{tsai2017providing, Kangasrasio2015, bruns2015should, zhao2010}, perceived usefulness \cite{tintarev2015inspection, Kangasrasio2015, chen2012cofeel, zhao2010}, user perception of transparency \cite{tsai2017providing, tsai2021effects}, trust \cite{schaffer2015hypothetical, bruns2015should, loepp2014choice}, user experience \cite{ schaffer2015hypothetical, knijnenburg2012inspectability, Donovan, 2012tasteweights, setfusion2014}, cognitive load and acceptance of recommendations \cite{jin2018effects}, user satisfaction \cite{tsai2017providing, jin2016go, linkedvis2013, smallworlds, talkexplorer} and user acceptance \cite{Kangasrasio2015, linkedvis2013}. These findings highlight the impact of user control on various recommendation goals, suggesting that these goals may interact with each other. However, to the best of our knowledge, no work has yet investigated the impact of user control on transparency, trust, satisfaction, and perceived quality together, nor has there been an investigation into how these goals relate to each other in an interactive recommendation context. Moreover, varying levels of user control can influence these recommendation goals differently, as each adjustment in the control mechanisms has direct effects on the recommendations \cite{harambam}. Furthermore, different users have different needs for control \cite{knijnenburg2011each,jin2018effects}. Despite this significance, different levels of user control in RSs have not been extensively explored, except only a few studies \cite{jin2017different, jin2018effects, harambam}. Moreover, only a few research works have attempted including control in the context of an ERS \cite{ooge23steering,bustos2020edurecomsys,BOUSBAHI20151813,zapata2015evaluation,valtolina2024design,michlik2010exercises,jung2019video}. More specifically in the context of ERSs, different levels of user control have not yet been implemented. Furthermore, no research has yet investigated the simultaneous effects of different levels of user control on transparency, trust, satisfaction, and perceived quality in an interactive or educational recommendation context.

This paper addresses these gaps by introducing user control at multiple levels within the ERS module of the MOOC platform CourseMapper \cite{Ain2022}. This paper extends our prior research on introducing different levels of control in an ERS \cite{ain2025designing} where we enable users to interact with the input (i.e., user profile), process (i.e., recommendation algorithm), and output (i.e., recommendations) of the ERS. Additionally, we provide a systematic design of the ERS, addressing this gap in the literature on both interactive and educational recommender systems.
Particularly in this paper, we conducted a between-subjects user study (N=184), using a mixed-methods evaluation approach for a deeper investigation of the impact of user control on users’ perceptions of the recommendation goals of transparency, trust, satisfaction, and perceived quality. Furthermore, we explored the interactions between these goals and analyzed how different levels of user control (i.e., basic, input, process, and output) impact them. The following research questions guide our investigation:
\begin{itemize}
    \item \textbf{RQ1}. How do different levels of user control in an ERS affect users’ perceived control, transparency, trust, satisfaction, and perceived quality of recommendations? 
    \item \textbf{RQ2}. What are the effects of user control on transparency of, trust in, satisfaction with, and perceived recommendation quality of the ERS, and how do these goals relate to each other?
\end{itemize}
Our analysis revealed three main findings: (1) related to the impact of providing user control in an ERS, our results indicate that control over the user profile alone is sufficient to foster a positive perception of the system, while additional control features may enhance these perceptions, they do not serve as a foundation in establishing them, (2) when examining the effects of different levels of user control on different recommendation goals, only perceived control was significantly affected by the level of control, with input control exerting the strongest impact, and (3) concerning the relation between different recommendation goals, we found that different levels of control influence transparency, trust, satisfaction, and perceived quality in distinct ways, and that these goals are strongly interconnected.

The remainder of this paper is structured as follows. We begin by reviewing two areas of related research, namely educational recommender systems (ERSs) and interactive recommender systems (IntRSs) in Section \ref{relatedwork}. Following this, we present the technical details of the ERS module of our MOOC platform, CourseMapper, in Section \ref{description}. Afterwards, we discuss the systematic design of different levels of control in the ERS in Section \ref{design}. In Section \ref{evaluation}, we provide an overview of the online user study, followed by the presentation of our results along with the discussion of our findings in Section \ref{results}.
Finally, in Section \ref{limitations}, we highlight the limitations, and then summarize the work and outline our future research plans in Section \ref{concl}.
\section{Background and Related Work}\label{relatedwork} 
\label{background}
This section discusses related work on the application of recommender systems in the educational domain and interactive recommender systems that support user interaction with and control of recommender systems.
\subsection{Educational Recommender Systems} \label{ERS}
Educational recommender systems (ERSs) play a crucial role in personalized learning environments by offering tailored recommendations that enhance the educational experience. These systems utilize a range of algorithms and data sources to suggest resources, courses, and activities aligned with individual learners' needs, preferences, and learning styles \cite{ain2024learner}.
ERSs have been facilitating learning and teaching in various ways \cite{drachsler2015panorama}. For example, to recommend learning materials to support instructors in online programming courses \cite{chau2018learning}, to recommend educational activities to a group \cite{fotopoulou2020interactive}, to recommend courses \cite{farzan2011encouraging, aher2013combination}, to suggest additional learning resources \cite{tang2005smart}, to support teachers to
improve their courses or monitor their teaching resources \cite{gallego2013enhanced, garcia2009architecture}, and to provide recommendations while preparing for the oral examination of a language learning course \cite{santos2016toward}. Most of these ERSs either propose algorithmic enhancements or new frameworks for recommendation, or implement existing/new recommendation techniques in an educational context,
such as collaborative filtering \cite{bustos2020edurecomsys}, emotion detection \cite{bustos2020edurecomsys}, content-based similarity \cite{BOUSBAHI20151813}, and data mining and machine learning \cite{fotopoulou2020interactive}. For an in-depth overview on ERSs, readers can refer to the recent literature reviews \cite{2024videosbasedRS, da2023systematic, khanal2020systematic}.

Despite extensive research on ERSs, little attention has been given to enhancing interactivity by offering more user control options in the user interfaces (UIs) of the ERS. Only a few attempts have been made to provide interactivity and control in ERSs.
\citet{ooge23steering} presented an interactive ERS that allows adolescents to initialize and adjust their mastery level in mathematics for exercise recommendation. Users can influence future recommendations through a control mechanism that lets them request easier or harder exercises after completing an exercise. In addition, the system provides a visualization showing how users’ interactions affect their estimated mastery level over time. \citet{bustos2020edurecomsys} presented an interactive ERS where the user can search for educational resources based on three main criteria, namely keywords, category, and type of resource. The user can view recommendations generated through collaborative filtering or emotion detection. In the list of recommendations, the user can view more details using 'details' button. Moreover, they can provide feedback to the recommendations using five-star rating as well as mark them as favorite to view them later. Furthermore, the user can write reviews about the recommended resources in the comments section.
\citet{BOUSBAHI20151813} proposed an interactive MOOC recommender, where users can interact with the system to formulate the request as an input to the RS, e.g., add keyword for course title (text input), and select features (e.g., course fee, availability, language) using checkboxes and dropdown.
\citet{zapata2015evaluation} presented a group recommender (DELPHOS) that recommends learning objects (LOs) to a group of individuals. As an input, similar to a search engine, the user defines the desired search parameters based on a required text query or keywords, some optional metadata values and different filtering or recommendation criteria using sliders and checkboxes. Afterwards, DELPHOS shows the user a ranked list of recommended LOs which users can rate on a Likert scale of five stars, group members can add one or more tags to LOs, as well as add personal comments or additional information to them. In the area of conversational RSs in education, \citet{valtolina2024design} presented an intelligent chatbot-based RS to assist teachers in their activities by suggesting the best LOs and how to combine them according to their prerequisites and outcomes. The interaction with the RS is via text input where chatbot asks specific questions and the user provide answers in textual format. 
The ERS in \cite{michlik2010exercises} enabled interaction by explicitly asking learners for feedback on exercises’ difficulty after completing them. The ERS presented in \cite{jung2019video} enables learners to influence video recommendations by adjusting system parameters such as the relative importance of content similarity, readability level, and topic tags. By manipulating these parameters, users can tailor future recommendation to their preferences, for example by prioritizing easier or more complex videos or focusing on specific topics.
While there are few attempts to make ERSs more transparent by introducing
open learner models (OLMs) (e.g., \cite{barria2019explaining, abdi2020complementing}), the used OLMs, however, just show learners their system-generated interests, but do not allow learners to interact with them or modify them. 

In conclusion, ERSs tend to offer less user control and interactivity compared to RSs in other domains such as e-commerce, entertainment, and social media. One reason for this may be the concern that increasing user control in ERSs could overwhelm learners and negatively impact the learning experience. 
Furthermore, most attempts to introduce interactivity in ERSs provide limited and fragmented user control, which is typically restricted to either the input (e.g., search queries, filters) or output level (e.g., ratings, feedback). Very few systems consider control over the recommendation process itself, and none of the reviewed works provide comprehensive control across all three levels, i.e. input, process, and output within a unified framework. Consequently, there is a significant research opportunity to enhance ERS transparency and user engagement by incorporating control mechanisms across all aspects of the system. Moreover, prior work in interactive ERS has predominantly focused on system-centric evaluation metrics, such as recommendation algorithm evaluation, or model effectiveness. While these are important, they overlook user-centric outcomes, particularly how different levels of interaction and control affect users’ perceptions. Specifically, there is a lack of investigation into whether providing control at different levels leads to improvements in perceived control, transparency, trust, satisfaction, or perceived recommendation quality. To address these gaps, this paper presents an interactive ERS in which we introduce different levels of user control by allowing users to interact with all the three parts of the ERS, namely input (i.e., user profile), process (i.e., recommendation algorithm), and output (i.e., recommendations), and evaluates its impact on key user-centric measures, including perceived control, transparency, trust, satisfaction, and perceived quality.
\subsection{Interactive Recommender Systems}
Traditional research on recommender systems (RSs) has concentrated on improving the accuracy of recommendations by developing new algorithms or integrating additional data sources into the recommendation process \cite{jannach2017user}. However, many studies have demonstrated that higher accuracy does not always enhance the user experience of the RS \cite{he2016interactive}. Consequently, recent research has shifted towards understanding how different interface elements and user characteristics impact the overall user experience with RSs. Furthermore, an effective RS should also take into account factors such as transparency, to ensure societal value and trust \cite{tintarev2015explaining, harambam}. This shift in focus from purely algorithmic improvements to enhancing user experience has led to the development of what are known as interactive recommender systems (IntRS), which emphasize user control and interactivity to achieve greater transparency in RS \cite{he2016interactive, jugovac2017interacting, jannach2017user}.

IntRSs offer visual and exploratory UIs, allowing users to inspect the recommendation process and control the system to receive better recommendations \cite{he2016interactive}. These interactive, visual, and exploratory UIs progressively guide users towards their objectives, enhance their understanding of the system's functionality, and ultimately contribute to transparency \cite{he2016interactive}.
IntRSs have been developed in various domains including movies \cite{Svonava2012, schaffer2015hypothetical, loepp2014choice, Donovan, schafer2002, maxwell2015}, music \cite{2012tasteweights, saito2011, jin2018effects}, news \cite{harambam}, publications \cite{setfusion2014, bruns2015should}, tweets \cite{tintarev2015inspection}, group recommenders \cite{chen2012cofeel}, social recommenders \cite{wong2011diversity, zhao2010, smallworlds}, conference recommenders \cite{talkexplorer}, and job recommenders \cite{Donovan}. For the readers interested in a more comprehensive understanding of IntRSs, we recommend consulting the thorough literature reviews available on this subject in \cite{jannach2017user, he2016interactive}.

IntRSs can roughly be grouped by the level they allow users to take control on, namely the RS input (i.e., user profile), process (i.e., recommendation algorithm), and/or output (i.e., recommendations) \cite{he2016interactive, jannach2017user}. Interaction with the input of the RS allows users to create or modify their interests as they want. This level of control is provided by either allowing users to add, delete, or re-rate items in their profile using various UI elements \cite{schaffer2015hypothetical, 2012tasteweights, linkedvis2013, bruns2015should, jin2016go, schafer2002, tintarev2015inspection, jin2018effects, harambam} or allowing them to visually interact with visualizations of their interest profile to modify them \cite{Kangasrasio2015, wong2011diversity, Donovan}. Users can control the process of the RS by either choosing the recommendation algorithm \cite{Ekstrand2015LettingUC, talkexplorer, harambam}, or by manipulating the algorithmic parameters \cite{Ekstrand2015LettingUC, maxwell2015, 2012tasteweights, setfusion2014, linkedvis2013, jin2018effects}, using the UI elements provided. Lastly, users can control the output of the recommender by providing feedback to the recommendations \cite{loepp2014choice, chen2012cofeel}, or ordering and sorting the recommendations as they want \cite{2012tasteweights, harambam, maxwell2015, jin2018effects, tintarev2015inspection, talkexplorer, smallworlds, zhao2010, wong2011diversity, bruns2015should}, based on the interactive elements provided in the UI.

In summary, IntRSs offer varying levels of user interaction and control at three different levels, namely input, process, and output. As summarized in Table \ref{tab:UI design}, only a few IntRSs support interaction at all three levels \cite{2012tasteweights, saito2011, jin2018effects, harambam}. Most of the IntRSs enable user interaction at the input level, allowing users to provide or adjust their preferences. A smaller number of IntRSs facilitate interaction with the process where users can adjust algorithmic parameters. Only few recommenders allow switching between different algorithms. At the output level, most systems provide users with the ability to sort recommendations, while fewer offer options to give feedback on the recommendations. With regard to UI design, only the study in \cite{harambam} focused on the systematic design of the different control mechanisms in the UI of their proposed IntRS. To address these gaps, in this work, we introduce an interactive ERS that extends interactivity and user control which is not commonly found in the educational domain. Furthermore, we present the systematic design of the ERS in the MOOC platform CourseMapper in which we provide control across all three levels, i.e., input, process, and output of the ERS. 

The impact of user control has been investigated in the literature on IntRS in various ways. Many researchers have studied the impact of user control on one or more recommendation goals, namely, perceived quality of recommendations \cite{tsai2021effects, jin2016go}, perceived accuracy of
recommendations \cite{schaffer2015hypothetical, Donovan, bruns2015should, 2012tasteweights, maxwell2015, smallworlds}, recommendation novelty \cite{loepp2014choice}, recommendation diversity \cite{wong2011diversity}, usability \cite{Kangasrasio2015, bruns2015should, zhao2010, tsai2017providing}, ease of use and playfulness \cite{chen2012cofeel}, perceived usefulness \cite{Kangasrasio2015, chen2012cofeel, tintarev2015inspection, zhao2010}, user-perceived transparency \cite{tsai2017providing, tsai2021effects}, trust in the RS \cite{schaffer2015hypothetical, bruns2015should, loepp2014choice, jin2016go}, user experience with the RS \cite{knijnenburg2012inspectability, Donovan, 2012tasteweights, setfusion2014}, cognitive load and recommendation acceptance \cite{jin2018effects}, confidence with the RS \cite{schafer2002, jin2016go}, user satisfaction with the RS \cite{schaffer2015hypothetical, linkedvis2013, smallworlds, talkexplorer, tsai2017providing, jin2016go}, behavioural intentions of the user \cite{jin2016go}, and user acceptance of the RS \cite{Kangasrasio2015, linkedvis2013, jin2016go}.
While trust and satisfaction were the focus of a considerable number of studies, transparency remains under-explored.
Moreover, there is a notable gap that no research has comprehensively studied the effects of user control on transparency, trust, satisfaction, and perceived quality together in the same study. Additionally, the impact of these goals on each other has yet to be explored in the interactive recommendation context. To address this research gap, in this paper, we study the impact of user control on transparency, trust, satisfaction, and perceived quality. Moreover, we investigate the relationships between these goals.

While user control has demonstrated a prominent impact on the accuracy and effectiveness of recommendations, in terms of different levels of user control, understanding the controllability needs of individual users is crucial while designing the control mechanisms in RS \cite{knijnenburg2011each}. Furthermore, different levels of control are required because not all users have the same need for cognitive load and different domain knowledge asks for different levels of control \cite{jin2018effects}. Therefore, combining multiple control components potentially increases acceptance without increasing cognitive load significantly \cite{jin2018effects}. 
Evaluating user control at different levels is still underexplored in IntRS literature with the exception of only a few \cite{jin2018effects, jin2017different, harambam}. \citet{jin2017different} categorized typical user controls into three levels (high, medium, and low) and examined how these varying levels of user control influence cognitive load and the quality of recommendations. The authors found that high control yields the best recommendations but increases cognitive load, with most users preferring low to medium control. 
In a similar vein, \citet{jin2018effects} studied the effects of three control components (output, input, process) on cognitive load and recommendation acceptance for two personal characteristics i.e. musical sophistication
and visual memory capacity. The results reveal that controlling the output (recommendations) is the most favorable single control element. In addition,
controlling user profile (input) and algorithm parameters (process) was the most beneficial setting with multiple controls. Moreover, the settings of user
control significantly influence cognitive load and recommenation acceptance. Additionally, \citet{harambam} conducted four focus groups, with 21 Dutch news readers, to systematically investigate how individuals assess various control mechanisms with the input, process, and output phases of a News Recommender Prototype (NRP). The findings indicated that a comprehensible user profile, combined with the ability to influence the recommendation algorithms, is highly valued by the users. Therefore, it can be argued that it is beneficial to combine multiple control components in a RS to achieve better recommendation acceptance and user experience. 

To address these gaps, in this work, we introduce an interactive ERS that extends interactivity and user control which is not commonly found in the educational domain. Concretely, we introduce control across all the three levels, i.e., input, process, and output of the ERS. Moreover, we investigate the effects of different levels of user control on three recommendation goals, namely transparency (i.e, explain how the system works) \cite{tintarev2015explaining, hellmann2022development}, trust (i.e., increase user's confidence in the system) \cite{mcknight2009trust}, satisfaction (i.e., increase the ease of use or enjoyment) \cite{knijnenburg2012experimental}, and perceived quality (i.e., how relevant and useful users find the recommendations) together in the same study. 

\begin{table}
\caption{Different levels of user control provided and evaluated in IntRS literature. Where \checkmark indicates the level of control provided or the recommendation goal evaluated.}
\label{controltable}
\centering
\resizebox{\textwidth}{!}{
\begin{tabular}{|>{\hspace{0pt}}m{0.19\linewidth}|>{\centering\hspace{0pt}}m{0.048\linewidth}|>{\centering\hspace{0pt}}m{0.048\linewidth}|>{\centering\hspace{0pt}}m{0.048\linewidth}||>{\hspace{0pt}}m{0.048\linewidth}|>{\hspace{0pt}}m{0.048\linewidth}|>{\hspace{0pt}}m{0.048\linewidth}|>{\centering\hspace{0pt}}m{0.048\linewidth}|>{\centering\hspace{0pt}}m{0.048\linewidth}|>{\centering\hspace{0pt}}m{0.048\linewidth}|>{\centering\hspace{0pt}}m{0.048\linewidth}|>{\centering\hspace{0pt}}m{0.048\linewidth}|>{\centering\hspace{0pt}}m{0.048\linewidth}|>{\centering\hspace{0pt}}m{0.048\linewidth}|>{\centering\hspace{0pt}}m{0.048\linewidth}|>{\centering\hspace{0pt}}m{0.048\linewidth}|>{\centering\hspace{0pt}}m{0.048\linewidth}|>{\centering\hspace{0pt}}m{0.048\linewidth}|>{\centering\arraybackslash\hspace{0pt}}m{0.06\linewidth}|} 
\hline
\multicolumn{1}{|>{\centering\hspace{0pt}}c|}{Paper} & \multicolumn{3}{|>{\centering\hspace{0pt}}c|}{Level of Control} & \multicolumn{15}{|>{\centering\arraybackslash\hspace{0pt}}c|}{Evaluation Metrics} \\ 
\hline
 & \rotatebox{90}{Input} & \rotatebox{90}{Process} & \rotatebox{90}{Output} & \rotatebox{90}{Perceived Quality} & \rotatebox{90}{Behavioral Intentions} & \rotatebox{90}{Control} & \rotatebox{90}{Trust} & \rotatebox{90}{Transparency} & \rotatebox{90}{Novelty} & \rotatebox{90}{Diversity} & \rotatebox{90}{Confidence} & \rotatebox{90}{Satisfaction} & \rotatebox{90}{Perceived accuracy} & \rotatebox{90}{User experience} & \rotatebox{90}{Usability} & \rotatebox{90}{Perceived usefulness} & \rotatebox{90}{User acceptance} & \rotatebox{90}{Cognitive load} \\ 
\hline
\citet{ooge23steering} & \checkmark &  &  &  &  &  & \cellcolor{yellow!25}\checkmark &  &  &  &  &  &  &  &  &  &  &  \\ \hline
\citet{bustos2020edurecomsys} & \checkmark &  & \checkmark &  &  &  &  &  &  &  &  &  &  &  &  &  &  &  \\ \hline
\citet{BOUSBAHI20151813} & \checkmark &  &  &  &  &  &  &  &  &  &  &  &  &  &  &  &  &  \\ \hline
\citet{zapata2015evaluation} & \checkmark &  & \checkmark &  &  &  &  &  &  &  &  &  &  &  &  &  &  &  \\ \hline
\citet{valtolina2024design} & \checkmark &  &  &  &  &  &  &  &  &  &  &  &  &  &  &  &  &  \\ \hline
\citet{michlik2010exercises} &  &  & \checkmark &  &  &  &  &  &  &  &  &  &  &  &  &  &  &  \\ \hline
\citet{jung2019video} &  & \checkmark  &  &  &  &  &  &  &  &  &  & \cellcolor{green!25}\checkmark &  &  &  &  &  &  \\ \hline
\citet{jin2016go} & \checkmark & \checkmark &  & \cellcolor{green!25}\checkmark & \cellcolor{green!25}\checkmark &  & \cellcolor{red!25}\checkmark &  &  &  &  & \cellcolor{red!25}\checkmark &  &  &  &  &  &  \\ \hline 
\citet{schafer2002} & \checkmark &  &  &  &  &  &  &  &  &  & \cellcolor{red!25}\checkmark &  &  &  &  &  &  &  \\ \hline 
\citet{schaffer2015hypothetical} & \checkmark &  &  &  &  &  & \cellcolor{green!25}\checkmark &  &  &  &  &  & \cellcolor{green!25}\checkmark & \cellcolor{green!25}\checkmark &  &  &  &  \\ \hline 
\citet{Kangasrasio2015} & \checkmark &  &  &  &  &  &  &  &  &  &  &  &  &  & \cellcolor{yellow!25}\checkmark & \cellcolor{yellow!25}\checkmark & \cellcolor{green!25}\checkmark &  \\ \hline 
\citet{Donovan} & \checkmark &  &  &  &  &  &  &  &  &  &  &  & \cellcolor{green!25}\checkmark & \cellcolor{green!25}\checkmark &  &  &  &  \\ \hline 
\citet{tsai2017providing} &  & \checkmark &  &  &  &  &  & \cellcolor{green!25}\checkmark &  &  &  & \cellcolor{green!25}\checkmark &  &  & \cellcolor{green!25}\checkmark &  &  &  \\ \hline
\citet{tsai2021effects} &  & \checkmark &  & \cellcolor{green!25}\checkmark &  &  &  & \cellcolor{green!25}\checkmark &  &  &  & \cellcolor{green!25}\checkmark &  &  &  &  &  &  \\ \hline
\citet{Ekstrand2015LettingUC} &  & \checkmark &  &  &  &  &  &  &  &  &  &  &  &  &  &  &  &  \\ \hline 
\citet{loepp2014choice} & \checkmark &  & \checkmark &  &  &  & \cellcolor{green!25}\checkmark &  & \cellcolor{green!25}\checkmark &  &  &  &  &  &  &  &  &  \\ \hline 
\citet{Svonava2012} &  &  & \checkmark &  &  &  &  &  &  &  &  &  &  &  &  &  &  &  \\ \hline 
\citet{maxwell2015} &  &  & \checkmark &  &  &  &  &  &  &  &  &  & \cellcolor{green!25}\checkmark &  &  &  &  &  \\ \hline 
\citet{zhao2010} &  &  & \checkmark &  &  &  &  &  &  &  &  &  &  &  & \cellcolor{green!25}\checkmark & \cellcolor{green!25}\checkmark &  &  \\ \hline 
\citet{smallworlds} &  &  & \checkmark &  &  &  &  &  &  &  &  & \cellcolor{green!25}\checkmark & \cellcolor{green!25}\checkmark &  &  &  &  &  \\ \hline 
\citet{chen2012cofeel} &  &  & \checkmark &  &  &  &  &  &  &  &  &  &  &  &  & \cellcolor{green!25}\checkmark &  &  \\ \hline 
\citet{tintarev2015inspection} & \checkmark &  & \checkmark &  &  & \cellcolor{green!25}\checkmark &  &  &  &  &  &  &  &  &  & \cellcolor{green!25}\checkmark &  &  \\ \hline 
\citet{wong2011diversity} & \checkmark &  & \checkmark &  &  &  &  &  &  & \cellcolor{green!25}\checkmark &  & \cellcolor{green!25}\checkmark &  &  &  &  &  &  \\ \hline 
\citet{setfusion2014}, \cite{parra2015user} &  & \checkmark & \checkmark &  &  &  &  &  &  &  &  &  &  & \cellcolor{green!25}\checkmark &  &  &  &  \\ \hline 
\citet{talkexplorer} &  & \checkmark & \checkmark &  &  &  &  &  &  &  &  & \cellcolor{green!25}\checkmark &  &  &  & \cellcolor{green!25}\checkmark &  &  \\ \hline 
\citet{linkedvis2013} & \checkmark & \checkmark &  &  &  &  &  &  &  &  &  & \cellcolor{green!25}\checkmark &  &  &  &  & \cellcolor{green!25}\checkmark &  \\ \hline 
\citet{bruns2015should} & \checkmark & \checkmark &  &  &  &  & \cellcolor{green!25}\checkmark &  &  &  &  &  &  &  & \cellcolor{green!25}\checkmark &  &  &  \\ \hline 
\citet{2012tasteweights} & \checkmark & \checkmark & \checkmark &  &  &  &  &  &  &  &  & \cellcolor{green!25}\checkmark & \cellcolor{green!25}\checkmark & \cellcolor{green!25}\checkmark &  &  &  &  \\ \hline 
\citet{saito2011} & \checkmark & \checkmark & \checkmark &  &  &  &  &  &  &  &  &  &  &  &  &  &  &  \\ \hline 
\citet{jin2017different}, \cite{jin2018effects} & \checkmark & \checkmark & \checkmark &  &  &  &  &  &  &  &  &  &  &  &  &  & \cellcolor{green!25}\checkmark & \cellcolor{green!25}\checkmark \\ \hline 
\citet{harambam} & \checkmark & \checkmark & \checkmark &  &  & \cellcolor{green!25}\checkmark &  &  &  &  &  &  &  &  &  &  &  &  \\ \hline 
\textit{ERS in CourseMapper}& \checkmark & \checkmark & \checkmark &  &  & \cellcolor{green!25}\checkmark & \cellcolor{green!25}\checkmark & \cellcolor{green!25}\checkmark &  &  &  & \cellcolor{green!25}\checkmark &  &  &  &  &  &  \\ \hline
\end{tabular}
}
    \caption*{Legend: \crule[green!25]{0.2cm}{0.2cm} positive effect \crule[yellow!25]{0.2cm}{0.2cm} mixed effect \crule[red!25]{0.2cm}{0.2cm} no effect of controllability on the evaluated metrics}
\end{table}

\section{System Description}\label{description}
In this section, we present the ERS module in the MOOC platform CourseMapper \cite{Ain2022}, providing an overview of it's functionality, with a particular focus on the underlying recommendation algorithms and their technical details.
\subsection{Recommendation Algorithms}\label{algos}
\begin{figure}[h]
\includegraphics[width=0.7\linewidth]{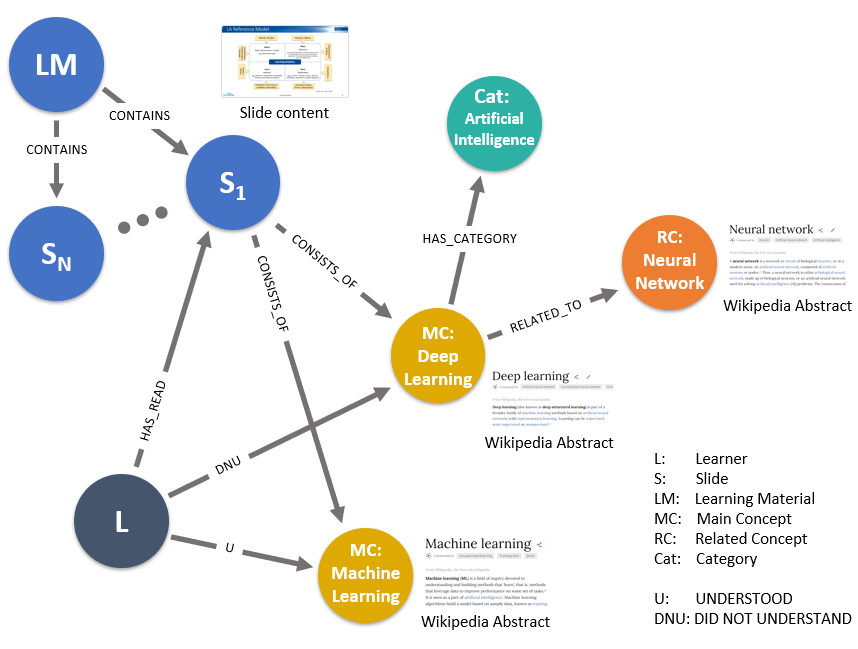}
\caption{A sample PKG for a learner.}
\label{pkg}
\end{figure}

Our ERS recommends learning resources (i.e., YouTube videos and Wikipedia articles) to the learners in the MOOC platform CourseMapper. Since for personalized recommendations, learners' individual knowledge state needs to be considered, we do that by creating a Personal Knowledge Graph (PKG) for each learner. The constructed PKG allows the ERS to provide personalized recommendations to the learners.
In CourseMapper UI, we provide a Did Not Understand (DNU) button at the bottom of each PDF learning material (lecture slides). When the learner clicks on this button, a slide-level knowledge graph (KG) is shown, consisting of the top-5 main concepts (MC) automatically extracted from the content of that slide \mbox{\cite{ain2023}}. After viewing the MCs of a particular slide the learner can select a MC and mark it as DNU or U (Understood). 
The slides which the learner has read and the MCs marked as U or DNU become nodes in the PKG, associated with that learner. In this way, the learners inform the system about their knowledge states themselves. A sample PKG for a learner is shown in Figure \ref{pkg}.  For the representation of entities (nodes) of the PKG, we use pre-trained transformer sentence encoders (SBERT) \cite{reimers2019sentence} to obtain a vector representation of each concept in the PKG by embedding its textual content on Wikipedia. Moreover, ConceptGCN \cite{alatrash2024conceptgcn} is used to enhance the representation of the nodes by aggregating the information of neighbors. Our ERS consists of four different recommendation algorithms that the user can select when generating recommendations.
Concretely, we implemented a PKG-based recommendation algorithm and a content-based recommendation algorithm, each with a keyphrase and a document variant.
\subsubsection{PKG-based recommendations:}
In this recommendation approach, we use the constructed PKG of the learner. First, the SBERT embedding of the Wikipedia content of DNU concepts of the learner are generated to represent the learner model. Then the DNU concepts of the learner are used to query YouTube and Wkipedia APIs to get candidates for recommendations. After that, from the title and description/abstract of the videos/articles, keyphrases are extracted, and SBERT is used to generate weighted average embedding of these keyphrases for the similarity computation in keyphrase variant. Whereas, for document variant, SBERT embedding of the title and description/abstract of the videos/articles is generated. Lastly, cosine similarities are computed between these embeddings and the representation of the learner model to recommend top-n candidates (videos and articles) to the learner.

\subsubsection{Content-based recommendations:}
In this recommendation approach, recommendations are generated based on the content of the learning material. First, SBERT embeddings of the MCs extracted from the slide are generated for the keyphrase variant. Whereas, SBERT embeddings of the of the whole content of the PDF slide are generated for the document variant. Then, top-n MCs extracted from the content of the slide are used to query the YouTube and Wkipedia APIs to get candidates for recommendations. After that, from the title and description/abstract of the videos/articles, keyphrases are extracted, and SBERT is used to generate weighted average embedding of these keyphrases for the similarity computation in keyphrase variant. Whereas, for document variant, SBERT embedding of the title and description/abstract of the videos/articles is generated. Lastly, cosine similarities are computed between these embeddings and the embedding of the MCs from the slide (keyphrase variant), or the embedding of the whole content of the slide (document variant). Lastly, the top-n most similar candidates (videos and articles) are recommended to the learner.
\begin{table} [h!]
  \caption{Factors ranking by users and their calculated default scores.}
  \label{tab:ranking}
  \resizebox{\textwidth}{!}{
  \begin{tabular}{lccccccc}
    \toprule
    Factors & Rank 1 & Rank 2 & Rank 3 & Rank 4 & Rank 5 & Rank 6 & Normalized score\\
    \midrule
    No. of likes on YouTube & 21 & 22 & 14 & 15 & 13 & 17 & 0.1774\\
    Creation date & 9 & 14 & 19 & 14 & 15 & 31 & 0.1415\\
    No. of views on YouTube & 22 & 22 & 16 & 18 & 16 & 8 & 0.1867\\
    Similarity score & 22 & 7 & 20 & 23 & 16 & 14 & 0.1690\\
    No. of saves on CourseMapper & 5 & 17 & 11 & 20 & 30 & 19 & 0.1391\\
    User rating on CourseMapper & 23 & 20 & 22 & 12 & 12 & 13 &0.1863\\
    \bottomrule
  \end{tabular}
  }
\end{table}
\subsection{Recommendations Ranking} \label{ranking}
The recommendations once generated need to be ranked in a defined order to present them to the learner. There are multiple factors in CourseMapper that contribute to the ranking of the recommendations. 
The factors influencing the ranking of recommendations are the different attributes of videos/articles recommended namely, 
number of likes on a video on YouTube, creation date of a video/article, number of views on a video on YouTube, similarity score of a video/article with user preferences, number of times a video is saved in CourseMapper, and user rating to a video/article in CourseMapper. 
Determining a default ranking based on these factors was a challenging task. Instead of assigning equal weights, we opted to let users prioritize the factors that mattered the most to them for ranking the recommendations. To gather their insights, we conducted an online survey using Google Forms, receiving 102 responses from 37 females and 65 males aged 18 to over 35 years. Most participants were familiar with online learning platforms (68\%) and recommender systems (75\%). The survey began with an overview of CourseMapper and the explanation of six ranking factors. Participants ranked the factors from 1 to 6 based on their perceived importance, as summarized in Table \ref{tab:ranking}. A Chi-square test confirmed a significant association between the factors and user-assigned ranks (p < 0.05). Consequently, 
we employed a direct-weighting technique to assign relative importance to each factor based on participants' responses \cite{singh2021review}. To transform the data into a proportional scale where the total always sums to 1, we applied Manhattan normalization. The obtained normalized scores (see Table \ref{tab:ranking}) ensure that the assigned weights are standardized and interpretable, allowing for direct comparison of factors and maintaining the relative importance of each factor. This approach provides a straightforward yet robust method to standardize and integrate multiple factors in the recommendation ranking process.
The normalized scores revealed that the participants ranked the factors in the following order: 1) number of views on a video on YouTube, 2) user rating of a video/article in CourseMapper, 3) number of likes on a video on YouTube, 4) similarity score of a video/article with user preferences, 5) creation date of a video/article, and 6) number of times a video is saved in CourseMapper.
\section{System Design} \label{design}
\begin{figure}[h]
  \centering
  \includegraphics[width=\linewidth]{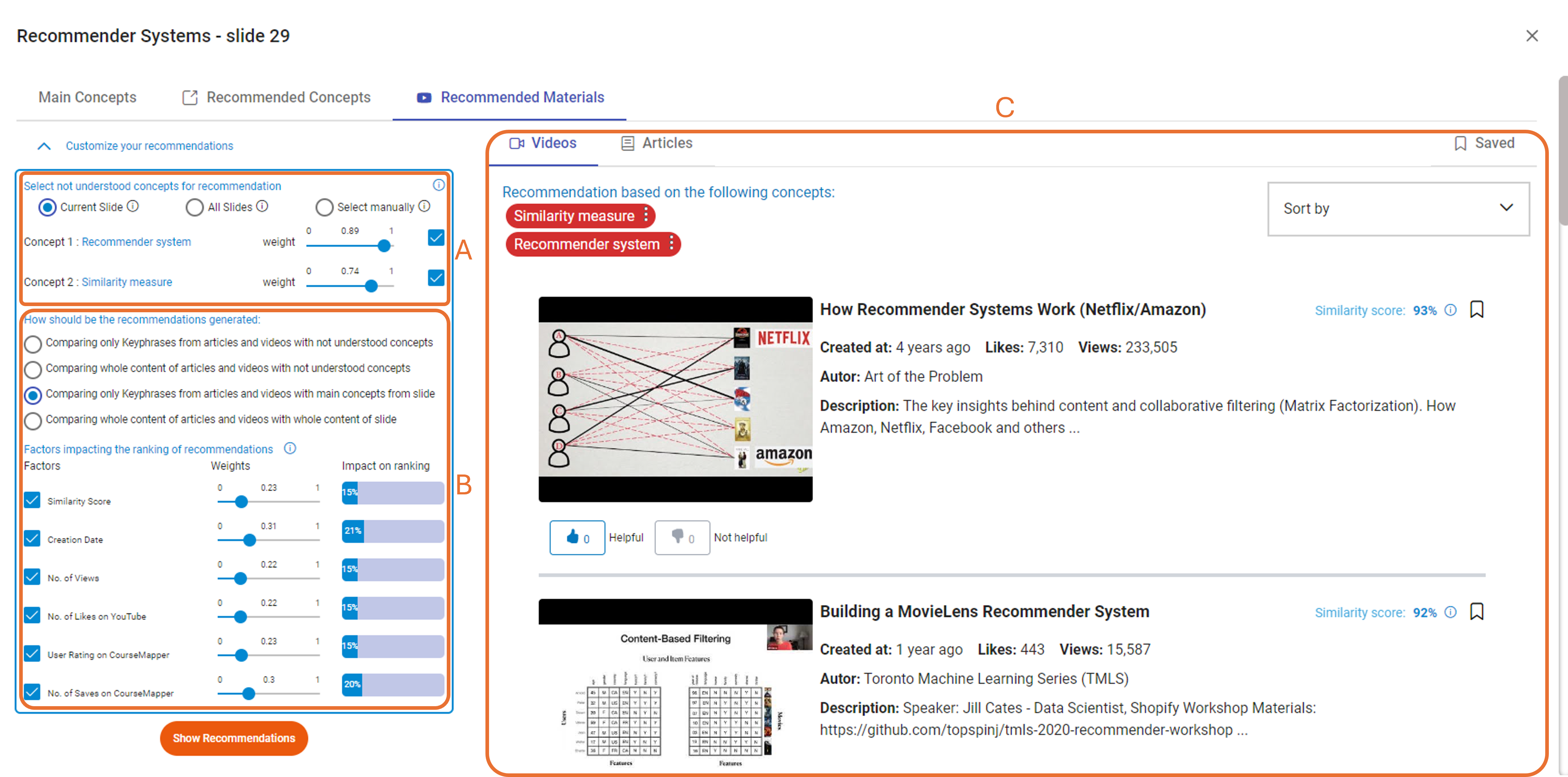}
  \caption{User Interface of the ERS in CourseMapper with three levels of user control: input (A), process (B), and output (C).}
  \label{overallinteraction}
\end{figure}

In this section, we present the design of the ERS module in the MOOC platform CourseMapper \cite{Ain2022}, which introduces user control at three different levels, namely input, process, and output (see Figure \ref{overallinteraction}). 
\subsection{User Interface Design}
In this section, we discuss the systematic approach taken to design interactive components for the UI of the ERS in CourseMapper, focusing on enhancing user control. We began by investigating the existing literature on IntRSs to identify a range of user control mechanisms and interaction options commonly employed in these systems, and analyzing their effectiveness in enhancing user control. The literature was explored focusing on answering the question: \textit{How user control has been added to the UI of the RSs to enable users to interact with different parts of the RS, i.e., input, process, and output?} Once the interaction mechanisms were identified, we chose the ones that are equally applicable to our context to ensure a better user experience. In this way, we designed interaction and control options in the UI of our ERS, focusing on the interaction with the input, process, and output of the ERS.

Beginning with the input of the IntRS, interaction with the input provides users the control to manage their preferences rather than the traditional method in which user preferences are estimated by observing their behavior over time. Many interaction options have been implemented at the input part of the recommenders to help users personalize their recommendations. Typically, to refine their requirements, users are asked to mark a set of items extracted by the system based on their past activity, using a binary scale in terms of “like/dislike” options. For instance, using ”Yes” or ”No” buttons \cite{saito2011}, or "Like" or "Dislike" buttons \cite{jin2016go}. Another option to interact with the input is provided by letting the users change the weights of their interests or re-rate items using sliders \cite{tintarev2015inspection, linkedvis2013, schaffer2015hypothetical, 2012tasteweights}, or using a pre-defined sliding scale ranging from ‘Strongly Disagree’ to ‘Strongly Agree’  \cite{wong2011diversity}. Another way is to let the users choose or modify their interests, for example, add or delete items in their profile using buttons with icons \cite{jin2018effects, schaffer2015hypothetical, linkedvis2013}, or radio buttons  \cite{tintarev2015inspection}, and re-rate items using sliders \cite{2012tasteweights, schaffer2015hypothetical, bruns2015should, harambam}. Alternatively, there are more complex methods for obtaining preferences. Examples include using filters for specific items \cite{bruns2015should}, drop down lists, checkboxes, and radio buttons to specify different dimensions of the interests \cite{jin2016go, schafer2002}, or using toggle buttons to enable/disable certain interests \cite{harambam}.
Furthermore, more advanced ways of interactions using visualizations are provided to the users using intent radar which the users can interact with using a mouse to move interest items \cite{Kangasrasio2015}, or interaction with a graph visualization of interests using a mouse to drag and drop items \cite{Donovan}.

Interaction with the process is commonly provided by allowing users to select or change the recommendation algorithm or tune the algorithm parameters. The selection of the algorithm is provided using radio buttons \cite{Ekstrand2015LettingUC}, text and icon based buttons \cite{harambam}, or selection using checkboxes \cite{talkexplorer}. To fine-tune the algorithm parameters multiple control options are provided to the users including radio buttons for feature selection \cite{saito2011}, buttons to add or remove parameters \cite{linkedvis2013}, and sliders to adjust weights of the parameters \cite{2012tasteweights, setfusion2014, bruns2015should, tsai2017providing, jin2018effects}.

Once the user’s preferences are identified, recommendation algorithm selected or modified, the system can provide tailored recommendations. Several control options and UI components have been proposed in the literature to interact with the output of the RS. Users can change the number of recommendations or filter the recommendations list using sliders \cite{Svonava2012, 2012tasteweights}, give feedback to the system using Yes/No \cite{saito2011, loepp2014choice} or Like/Dislike \cite{jin2018effects} buttons, give feedback about recommendations using buttons of different sizes referring to intensity of emotions \cite{chen2012cofeel}, or can give feedback using radio buttons \cite{maxwell2015}. Users are provided with the options to sort the recommendation list based on multiple options using buttons \cite{jin2018effects, tintarev2015inspection}, remove recommendations from the list using remove icon \cite{jin2018effects}, sort recommendations using drag and drop \cite{jin2018effects}, and reorder recommendations using toggles and sliders \cite{harambam}. Moreover, advanced interaction options are provided to the users when recommendations are presented visually instead of lists. The users can interact with the Venn diagram to examine and filter the recommended items \cite{setfusion2014}, explore the opinion space using mouse interaction to increase or decrease diameter of circular opinion space \cite{wong2011diversity}, mouse interaction with word cloud \cite{zhao2010}, drag and drop nodes in a graph \cite{smallworlds}, and arrange items in a clustermap using mouse drag interaction \cite{talkexplorer}.

\begin{figure*}[h!]
     \centering
     \begin{subfigure}{0.7\textwidth}
        
         \includegraphics[width=\linewidth]{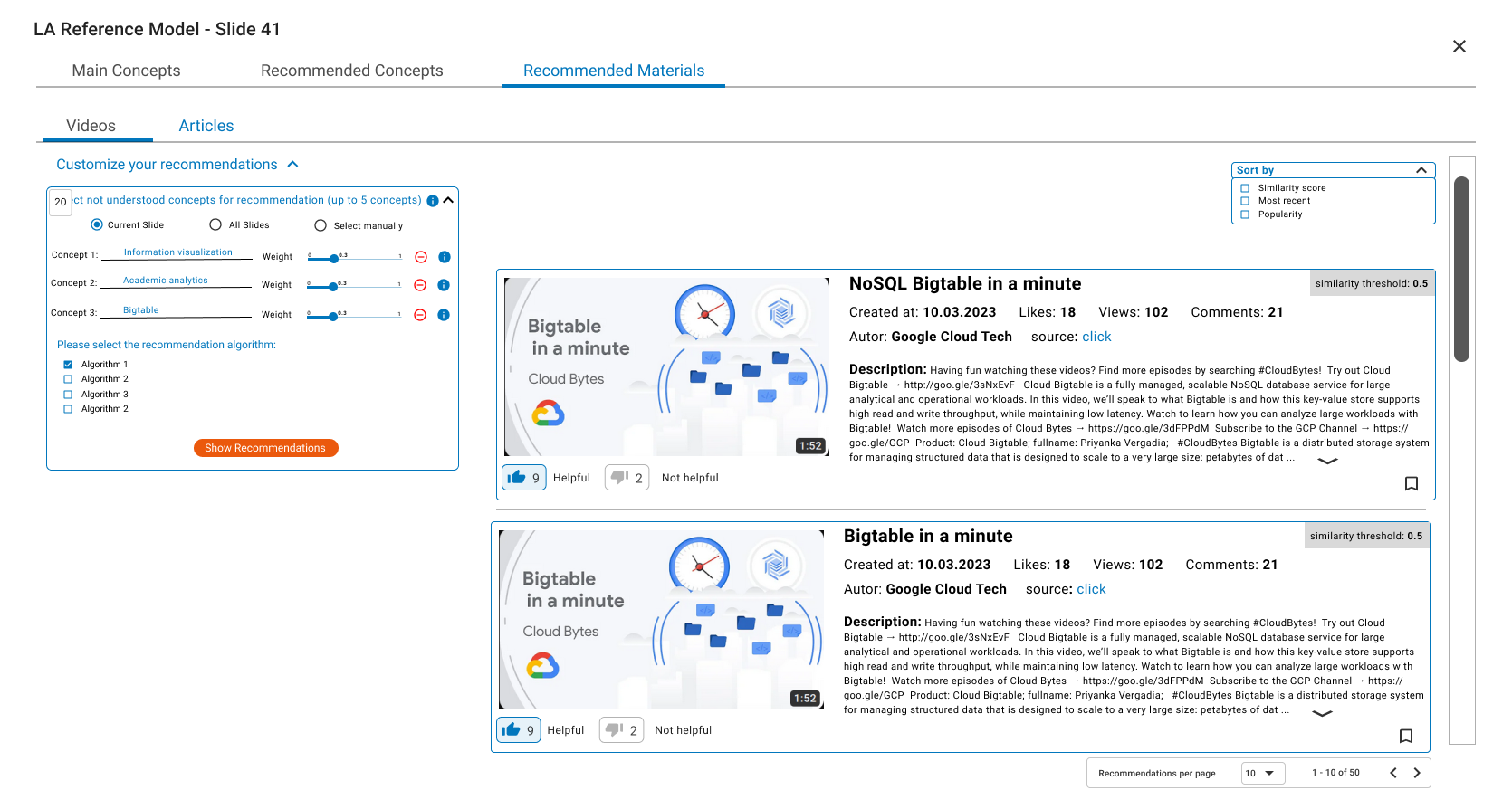}
         \caption{Initial prototype}
         \label{fig:initial}
     \end{subfigure}
     \hfill
     \begin{subfigure}{0.7\textwidth}
         \includegraphics[width=\linewidth]{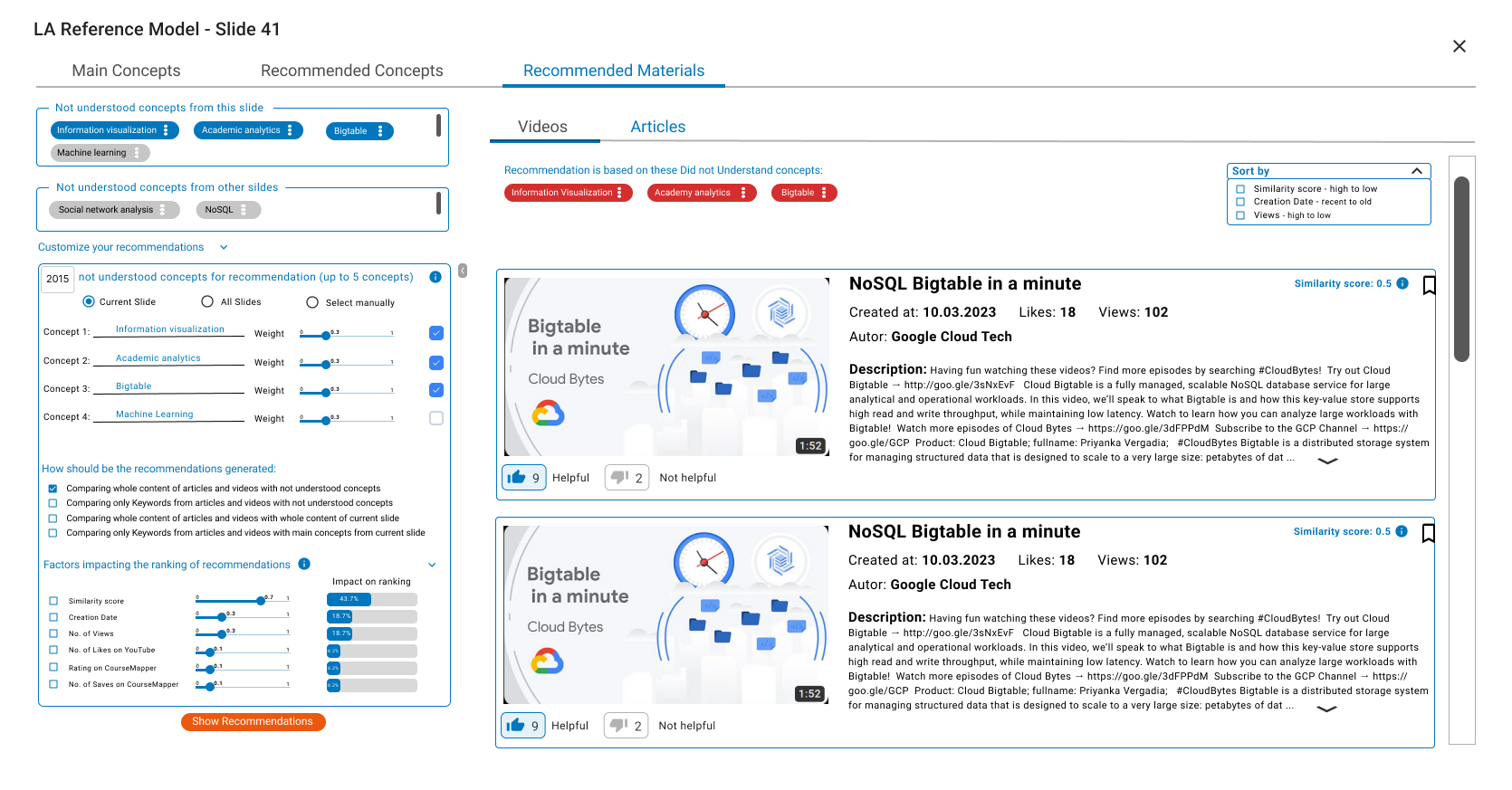}
       \caption{Improved prototype}
        \label{fig:final}
     \end{subfigure}   
        \caption{Prototypes for different levels of user control in the ERS.}
        \label{fig:prototypes}
\end{figure*}

From this analysis, we identified and selected the most widely adopted interaction techniques that align with our specific context and features, ensuring that our UI design is both intuitive and effective for our users (see Table \ref{tab:UI design}). After that, we started with the design of our ERS UI. Based on the UI elements identified to interact with the input, process, and output, we created initial prototypes (see Figure \ref{fig:initial}). The prototypes were discussed with the authors' team and different UI elements were refined and improved. For example, deciding the colors, optimal labels for buttons, and labels to represent the algorithms, providing UI for ranking the recommendations, and deciding whether or not to show the impact of ranking in progress bars. The improved prototypes (see Figure \ref{fig:final}) were then translated to the final design of the system, presented in the next sections.

\begin{table}[h]
  \caption{UI design elements for interaction with different components of the RS identified from literature.}
  \label{tab:UI design}
  \resizebox{\textwidth}{!}{%
    \begin{tabular}{lp{4cm}p{4cm}p{5cm}} 
      \toprule
      Paper & User control with input & User control with process & User control with output\\
      \midrule
      \citet{ooge23steering} & Slider to adjust mastery level & - & - \\
      \citet{bustos2020edurecomsys} & Search using keywords, category,  and type of resource & - & Feedback using 5-star rating, \newline mark as favorite button to view later, write reviews\\
      \citet{BOUSBAHI20151813} & Search keyword (text input), select features using checkboxes and dropdown & - & -\\
      \citet{zapata2015evaluation} & Search keyword (text input), Filtering criteria (sliders, checkboxes) & - & Rate recommendations (5-star Likert scale), write comments\\
      \citet{valtolina2024design} & text input (chat-bot based) & - & - \\
      \citet{michlik2010exercises} & - & - & Feedback on exercise difficulty using buttons\\
      \citet{jung2019video} & - & Sliders to adjust parameters weights & -\\
      \citet{jin2016go} & Like/Dislike buttons, \newline Selection using radio buttons, and dropdowns & - & - \\
      \citet{schafer2002} & Selection using radio buttons, checkboxes, and dropdowns & - & - \\
      \citet{schaffer2015hypothetical} & Add or delete buttons, \newline Sliders to re-rate items & - & - \\
      \citet{Kangasrasio2015} & Mouse interaction with radar & - & - \\
      \citet{Donovan} & Mouse interaction with graph & - & - \\
      \citet{linkedvis2013} & Sliders to adjust weights & Buttons to add or remove parameters & - \\
      \citet{bruns2015should} & Filter items using buttons & Sliders to adjust weights & - \\
      \citet{2012tasteweights} & Sliders to adjust weights & Sliders to adjust weights & Filter using sliders \\
      \citet{saito2011} & Yes/No buttons & Feature selection using radio buttons & Feedback using Yes/No buttons \\
      \citet{loepp2014choice} & - & - & Feedback using Yes/No buttons \\
      \citet{Svonava2012} & - & - & Filter using sliders \\
      \citet{jin2018effects} & Select input using buttons & Sliders to adjust weights & Feedback using Like/Dislike buttons, drag to sort, remove item \\
      \citet{maxwell2015} & - & - & Feedback using radio buttons \\
      \citet{chen2012cofeel} & - & - & Feedback using size of buttons \\
      \citet{harambam} & Toggle to enable/disable interests & Select using text buttons & Reorder using toggle \\
      \citet{tintarev2015inspection} & Sliders to adjust weights & - & Sort using buttons \\
      \citet{setfusion2014} & - & Sliders to adjust weights & Filter in Venn diagram \\
      \citet{wong2011diversity} & Fixed sliders to adjust weights & - & Mouse interaction with circular opinion space \\
      \citet{zhao2010} & - & - & Mouse interaction with word cloud \\
      \citet{smallworlds} & - & - & Drag and drop nodes in a graph \\
      \citet{talkexplorer} & - & Select using checkboxes & Drag and drop in cluster map \\
      \citet{tsai2017providing} & - & Sliders to adjust weights & - \\
      \citet{Ekstrand2015LettingUC} & - & Select using radio buttons & - \\
      \textbf{ERS in CourseMapper} & \textbf{Add, delete, select using buttons, include/exclude using checkboxes, adjust the weights using sliders} & \textbf{Select using radio buttons, adjust weights for ranking using sliders, view impact of adjustments in progress bars} & \textbf{Save using buttons, Sort using dropdown options, Feedback with connected input selection using Helpful/Not Helpful button}\\
      \bottomrule
    \end{tabular}
  }%
\end{table}

\subsection{User Control with the Input} \label{inputsection}
\begin{figure*}[!ht]
     \centering
         \begin{subfigure}{0.3\textwidth}
         \centering
         \includegraphics[width=\textwidth]{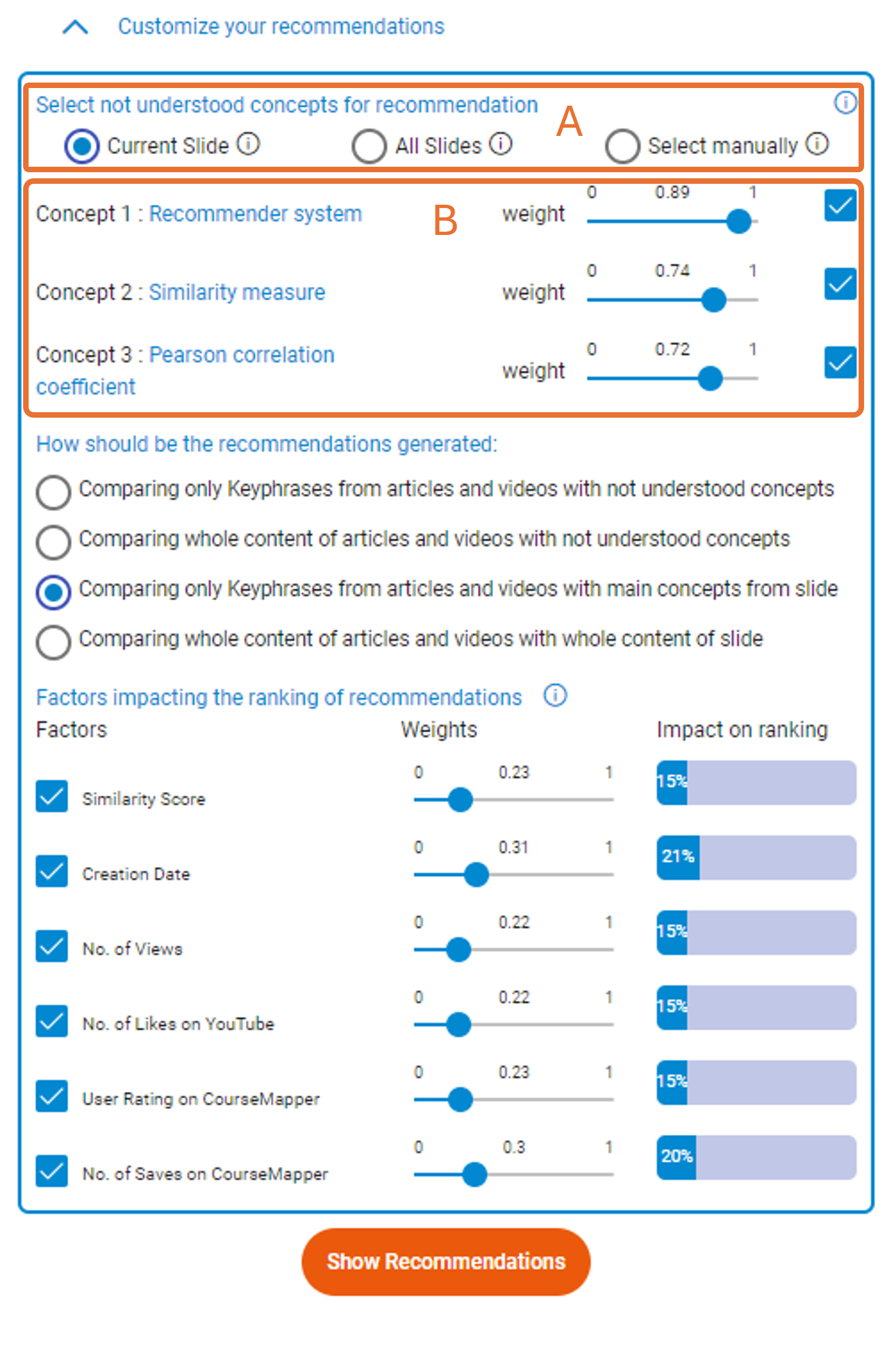}
         \caption{Select DNU concepts from the current slide as input for recommendation}
         \label{fig:current}
        \end{subfigure}
     \hfill
        \begin{subfigure}{0.3\textwidth}
         \centering
         \includegraphics[width=\textwidth]{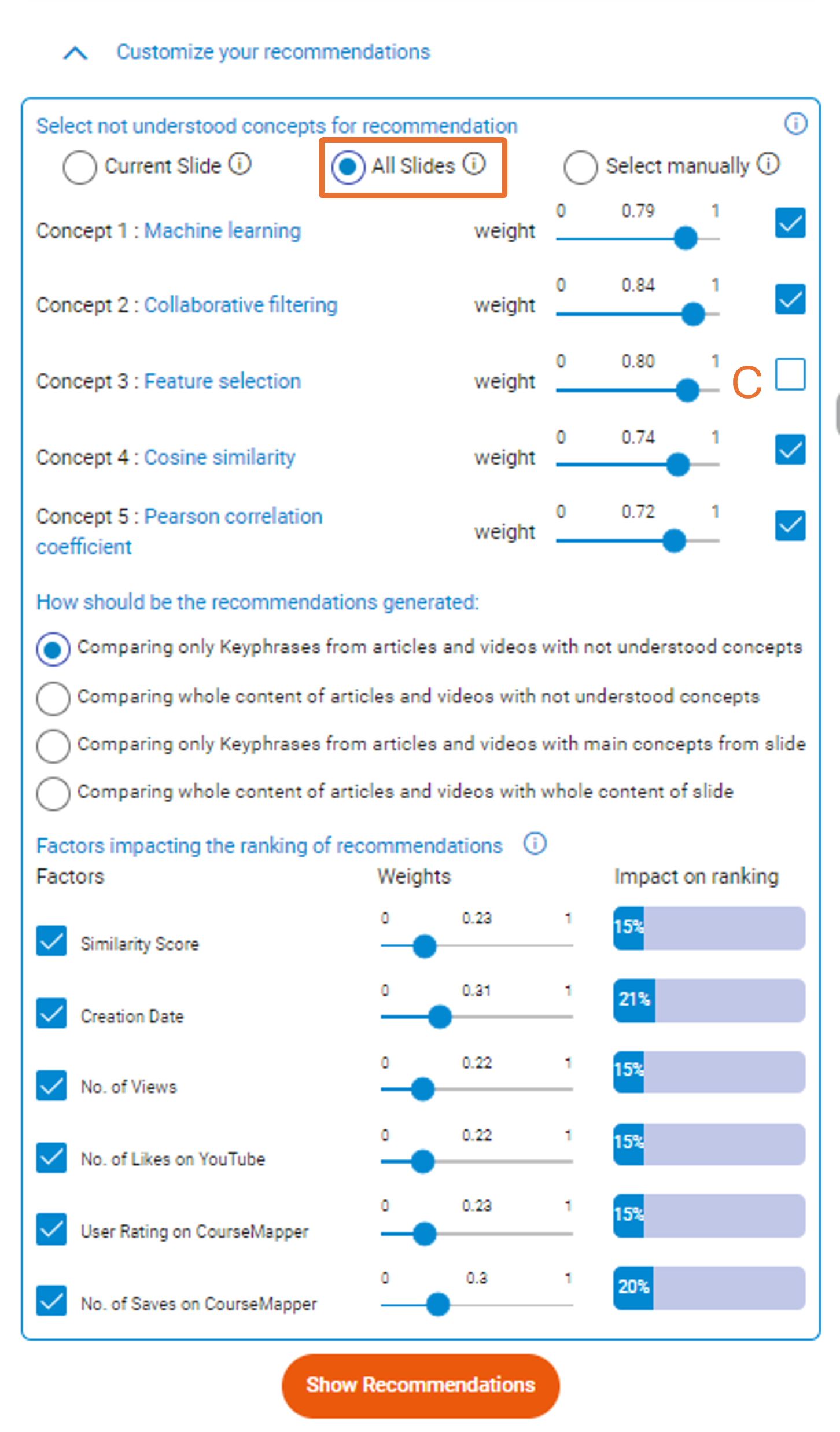}
       \caption{Select DNU concepts from all slides as input for recommendation}
        \label{fig:allslides}
        \end{subfigure}
          \hfill
         \begin{subfigure}{0.3\textwidth}
         \centering
         \includegraphics[width=\textwidth]{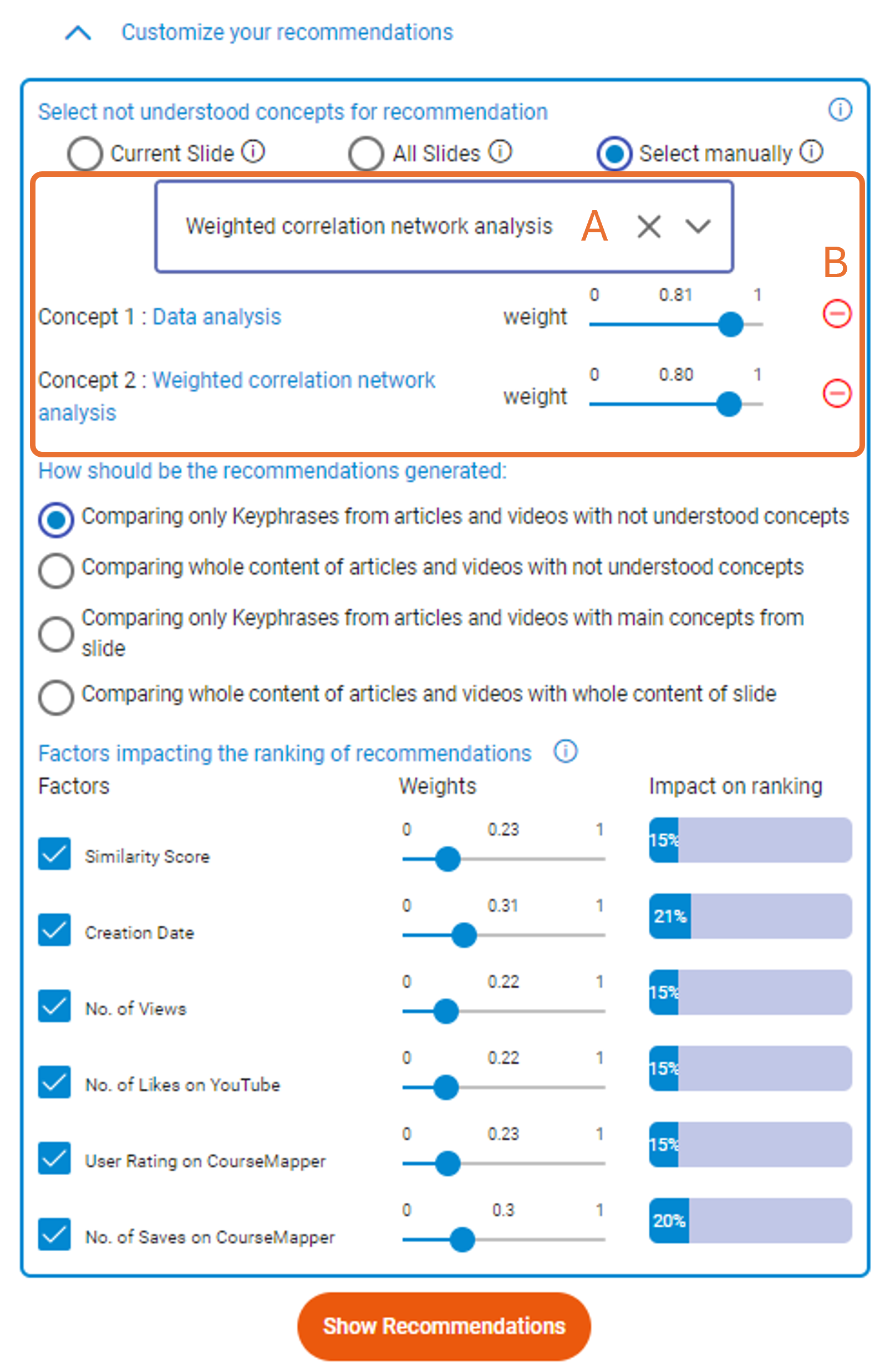}
       \caption{Manually select concepts from the learning material as input for recommendation}
        \label{fig:manually}
         \end{subfigure}
        \caption{User control with the input of the ERS.}
        \label{fig:input}
\end{figure*}

In the context of recommending learning resources, it is crucial to recommend accurate resources tailored to learners’ needs for
better learning outcomes \cite{ain2024learner}. Therefore, providing interaction around the input of the ERS will facilitate learners to directly communicate their interests to the system. A simple way to give users more control is to allow them to explicitly specify their interests and preferences rather than relying on the system to determine these preferences from their past interactions \cite{jannach2017user}. 
To facilitate interaction with the input, we provide control to the users to choose their interests and preferences explicitly to construct their learner model based on DNU concepts. 
Figure \ref{fig:input} shows the UI of the ERS with multiple options to interact with the input. First, the users can decide which DNU concepts they want the recommendations for, using the radio buttons to select 'Current Slide', 'All Slides', or 'Select manually' options (Figure \ref{fig:current} (A)). Clicking on 'Current Slide' and 'All Slides' retrieves the DNU concepts related to the current slide or all slides of the learning material, respectively (Figure \ref{fig:current} (B), \ref{fig:allslides}). Whereas, clicking on the 'Select manually' option let the user choose any concepts from the whole learning material using the dropdown list with a search option (Figure \ref{fig:manually} (A)). Once the DNUs are selected as input for the recommender, in all the three options, the user is provided the opportunity to manipulate the weights of the concepts based on their preferences if they find some concepts more important than others (Figure \ref{fig:current} (B)). The recommendations will be impacted by the changes in weights correspondingly. Furthermore, the user can include/exclude certain DNUs using checkboxes (Figure \ref{fig:allslides} (C)), as well as remove DNUs from the input list using '-' icon (Figure \ref{fig:manually} (B)).
\begin{figure*}[!ht]
    \centering
    \begin{subfigure}{0.32\textwidth}
        \includegraphics[width=\linewidth]{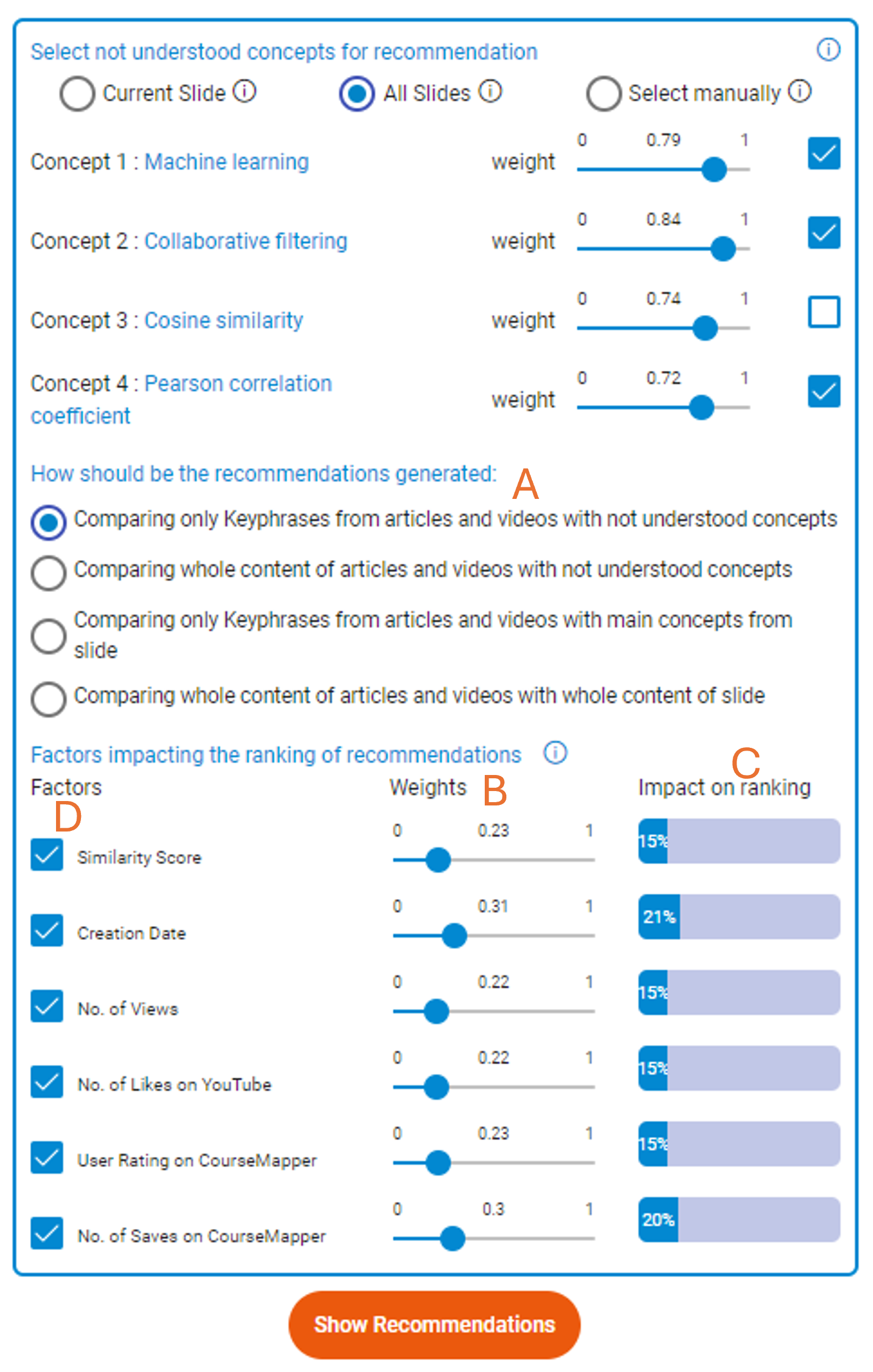}
        \caption{User control with the process of the ERS}
        \label{fig:process}
    \end{subfigure}
    \hfill
    \begin{subfigure}{0.67\textwidth}
        \raisebox{15mm}{ 
            \includegraphics[width=\linewidth]{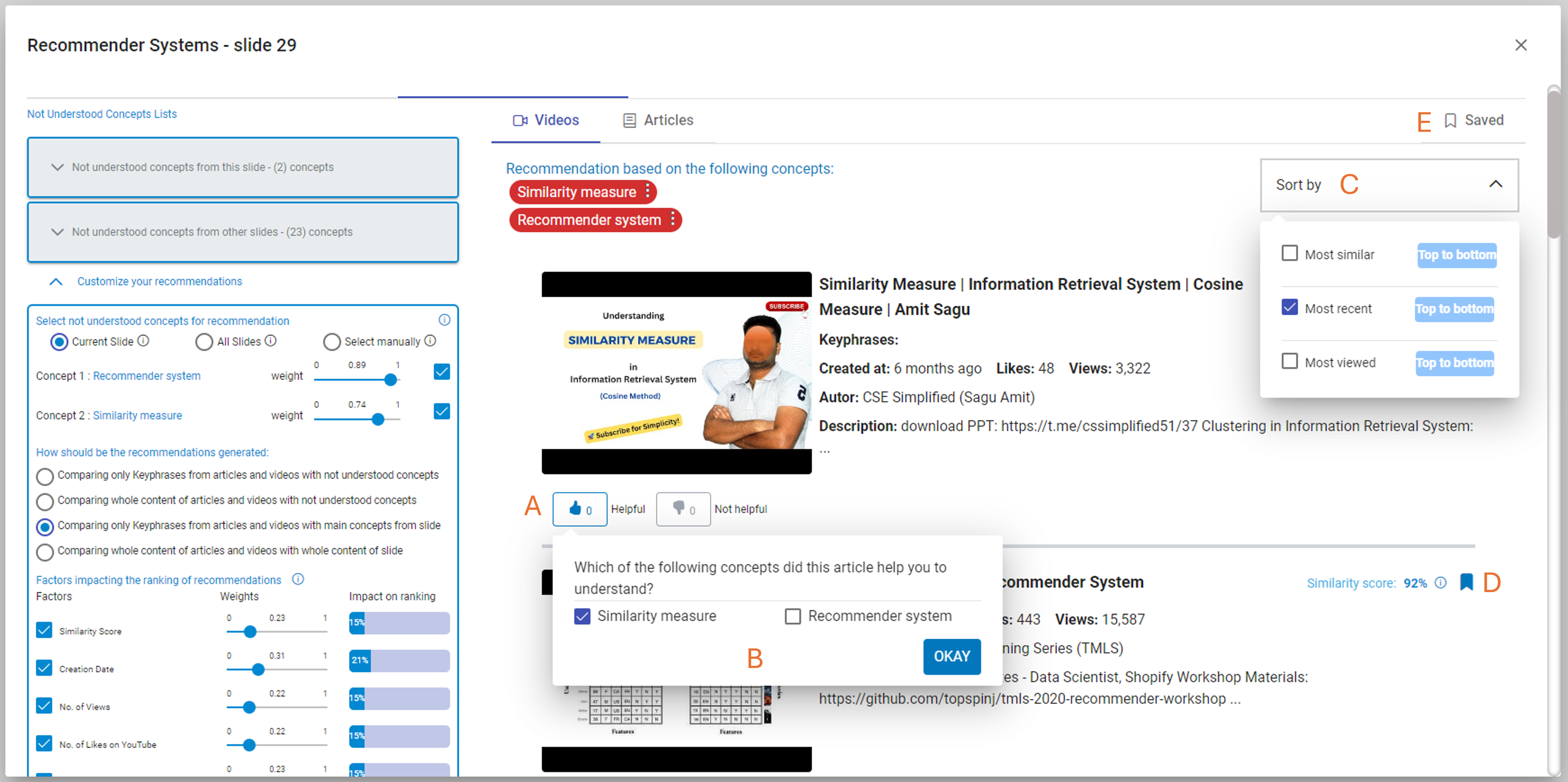}
        }
        \caption{User control with the output of the ERS}
        \label{fig:output}
    \end{subfigure}   
    \caption{User control with the process and output of the ERS.}
    \label{fig:prout}
\end{figure*}

\subsection{User Control with the Process} \label{processsection}
Interacting with the recommendation process allows users to choose or influence the recommendation strategy or algorithm parameters \cite{jannach2019explanations}. 
To introduce user control with the process of our ERS, we provide two options to the user. The first is to let them choose between the four recommendation algorithms (described in Section \ref{algos}) using radio buttons (Figure \ref{fig:process}, (A)). 
The options to select the algorithm are presented as an answer to the question "How should the recommendations be generated?", with four possible answers as: 1) Comparing only keyphrases from articles and videos with DNU concepts, 2) Comparing the whole content of articles and videos with DNU concepts, 3) Comparing only keyphrases from articles and videos with main concepts of the current slide, and 4) Comparing the whole content of articles and videos with whole content of the current slide. Once the user selects the algorithm, the second control option is to decide how they want the recommendations to be ranked. The user can view multiple factors impacting the ranking of the recommendations with their default weights (refer to Section \ref{ranking}). The users can adjust the weights of the factors using sliders (Figure \ref{fig:process}, (B)), where the impact of the weight change on the ranking is displayed in real time on the progress bar adjacent to each factor (Figure \ref{fig:process}, (C)). Furthermore, the user can select/de-select factors from the list using checkboxes if they do not want to include a factor in ranking (Figure \ref{fig:process}, (D)). 
\subsection{User Control with the Output} \label{outputsection}
We provide multiple control options to the user to interact with the output of the ERS (see Figure \ref{fig:output}). The first is to let users provide feedback to the system about the recommendations provided. To facilitate this, we provide a 'Helpful/Not Helpful' button with each recommended video or article (Figure \ref{fig:output}, (A)). Once the user clicks on 'Helpful' a dropdown menu appears which prompts the user to select the DNU concepts that the recommended video or article helped them understand (Figure \ref{fig:output}, (B)). This extended level of control enables the user to provide detailed feedback to the ERS, which can be used to improve future recommendations. Furthermore, the user can sort the recommendations using the sort option which opens a dropdown with multiple options (Figure \ref{fig:output}, (C)). The user can sort the recommendations based on their similarity score, creation date, or number of views. 
The third interaction option is to save the recommendations using a save icon (Figure \ref{fig:output}, (D)). These recommendations are saved to access them later in the 'Saved' tab (Figure \ref{fig:output}, (E)).
\section{User Study} \label{evaluation}
To evaluate the different levels of control introduced in the ERS, we conducted a between-subjects online user study using a mixed-method evaluation approach that combines quantitative and qualitative methods.
\subsection{Participants}
We recruited 184 participants (101 males, 83 females, mean age 31.66 and range between 18 and 45) through the Prolific crowd-sourcing platform \cite{prolific}. Participants were located in several countries, with the majority residing in the United Kingdom (N = 89, 48.1\%) and the United States (N = 74, 40.0\%). Smaller groups came from South Africa (N = 8, 4.3\%), Canada (N = 7, 3.8\%), Australia (N = 3, 1.6\%), Scotland (N = 2, 1.1\%), and Italy (N = 1, 0.5\%). Regarding prior experience, participants reported a moderate familiarity with recommender systems (M = 3.41, SD = 1.15) and relatively high familiarity with online learning systems (M = 4.00, SD = 0.90) on a 5-point Likert scale. We restricted the execution of the task to users with a task completion rate greater than 95\%, and number of tasks completed greater than 500. To ensure quality of responses a completion code was shown to participants only if they pass all the attention checks in the study and answer all the required questions. Moreover, as a proof of task completion the participants were required to upload the screenshot of the recommendation interface before starting the survey. The responses of 16 subjects were discarded due to this quality check (from an initial number of 200 participants), so only the responses of 184 subjects were used for the analysis. Before conducting the study, the power analysis suggested N=45 participants for each condition in a between-subjects study design for a statistical power of 0.80, $\alpha$=0.05 \cite{cohen2013statistical}). Participants were rewarded with £4.5 for a 30 minutes study based on Prolific's fixed standard rate of £9 per hour. The average time taken by the participants to complete the study was 29 minutes. 
\subsection{Questionnaires}
The goal of our study is to investigate the impact of user control in an ERS on users’ perceptions of control, transparency,
trust, satisfaction, and perceived recommendation quality.
\textit{Perceived control} was measured using 4 items adopted from \citet{knijnenburg2012experimental}, e.g., “I had limited control over the way CourseMapper made recommendations (reversed)”, the scale demonstrated good internal consistency in our sample (Cronbach’s $\alpha$ = .87). \textit{Transparency} was measured using 12 items adopted from \citet{hellmann2022development}, e.g., "The system provided information to understand why the videos were recommended", the scale demonstrated excellent internal consistency in our sample (Cronbach’s $\alpha$ = .90). \textit{Trust} was evaluated using the items adopted from \citet{mcknight2009trust}, e.g., "I can always rely on video recommendations in CourseMapper to understand a difficult concept", the scale demonstrated excellent internal consistency (Cronbach’s $\alpha$ = .96). To evaluate \textit{satisfaction}, we used 7 items adopted from \citet{knijnenburg2012experimental}, e.g., "I would recommend CourseMapper to others", the scale demonstrated excellent internal consistency (Cronbach’s $\alpha$ = .91). \textit{Perceived recommendation quality} was evaluated using 6 items adopted from \citet{knijnenburg2012experimental}, e.g., "The recommended videos were relevant", the scale demonstrated excellent internal consistency in our sample (Cronbach’s $\alpha$ = .92). In addition, we also adopted two items from \citet{knijnenburg2012experimental} to evaluate users' familiarity with recommender systems, e.g., "I am familiar with online recommender systems", the scale demonstrated excellent internal consistency (Cronbach’s $\alpha$ = .91). Furthermore, we used self-created items to measure users' familiarity with interacting with recommender systems and their familiarity with online learning platforms and MOOCS. All items except trust were measured with a 1–5 Likert-scale (1: Strongly disagree, 5: Strongly agree), whereas trust was measured using 1-7 Likert scale (1: Strongly disagree, 7: Strongly agree).
\subsection{Procedure}
The study follows a between-subjects design, and each participant was assigned randomly to one of the four conditions that represent the level of control with the ERS: \textit{basic control}, \textit{input control}, \textit{process control}, and \textit{output control}. The \textit{basic control} condition allows participants to mark their DNU concepts and generate recommendations (see Section \ref{algos}). No additional controls over the input, process, or output of the ERS are provided in this condition. Marking concepts as DNU constitutes the baseline interaction with the ERS, upon which the other control conditions build by introducing additional control features.
Therefore, the \textit{basic control} condition already provides a minimal level of interaction, and differences between conditions should be interpreted as resulting from the additional control features. The \textit{input control} condition enables more fine-grained control over the input (see Section \ref{inputsection}), the \textit{process control} condition provides mechanisms to influence the recommendation process (see Section \ref{processsection}), and the \textit{output control} condition allows users to interact with and adjust the recommended items (see Section \ref{outputsection}).
At the beginning of the study, participants provided informed consent by agreeing to the conditions outlined in the consent form, including voluntary participation and data confidentiality. They were also asked to provide their Prolific IDs. To avoid duplicate participation, IDs were automatically checked, and individuals who had already participated were screened out.
Participants then completed a short demographic survey capturing information on their location, age, familiarity with online learning platforms and MOOCs, as well as their familiarity with recommender systems (measured on a 5-point Likert scale).
Next, participants viewed a short demo video introducing CourseMapper and its core functionalities. The video showed how courses are structured into topics and channels and how users can access the learning materials. Moreover, the features related to marking concepts as not understood and generating recommendations were shown. Furthermore, in the recommendation step, the demo showed only the level of control to which the participant had been randomly assigned to. 
Following the demo, participants were required to sign up for CourseMapper and enroll in one of three available courses of their choice, namely Introduction to Python, Food and Nutrition, or Basic English Language Skills. The option to select among courses was provided to allow participants to engage with content aligned with their interests or prior knowledge. This was important to ensure meaningful interaction with the system, particularly for identifying concepts they do not understand (DNU) and evaluating the recommendations. The enrollment count in the three courses was as follows: 97 users in Food and Nutrition, 49 users in Python for Beginners, and 38 users in Basic English Language Skills. Once enrolled, the participants were assigned a task corresponding to their assigned level of control, as described below:
\begin{enumerate}[label=\arabic*.]
   \item \textbf{\textit{Basic control} condition:}
   Participants first enrolled in a course of their choice and accessed the provided PDF learning material. While engaging with the material, they identified concepts they found difficult to understand and marked them as not understood (DNU concepts). Based on these concepts, the system automatically generated a list of YouTube video recommendations. In this condition, participants had no control over the provided recommendations. They reviewed the recommended videos and then proceeded to the post-task survey, which captured their experiences with the system.
   \item \textbf{\textit{Input control} condition:}
   In this condition, participants exercised control over the input to the recommender. They could specify and revise the set of DNU concepts, modify the weights of the concepts using sliders, and exclude or include selected concepts using check boxes. They could generate an initial recommendation list and subsequently regenerate recommendations after adjusting the input set or weights. Participants then reviewed the recommended videos and completed the post-task survey regarding their experience.
    \item \textbf{\textit{Process control} condition:}
   In this condition, participants exercised control over the recommendation process. They could select among multiple recommendation algorithms using radio buttons and adjust the post-generation ranking method according to their preferences using sliders and check boxes. After generating recommendations based on their chosen settings, participants reviewed the recommended videos and then completed the post-task survey about their experience.
   \item \textbf{\textit{Output control} condition:}
   In this condition, participants exercised control over the output presentation and management of recommendations. They could re-order the list (e.g., sort by recency to view the newest videos first), save/bookmark selected videos for later access, and provide item-level feedback on the usefulness of recommended videos. After configuring and reviewing the recommendations, participants completed the post-task survey about their experience with the system.
\end{enumerate}
It is important to clarify here that participants interacted only with condition-specific interface elements based on the assigned task, ensuring that the interaction targeted only one control aspect (i.e. basic, input, process, or output) for each participant. After completing the assigned task, participants proceeded to the survey designed to capture their perceptions of using the ERS based on the questionnaire items described earlier. The survey was implemented using the SoSci Survey tool \cite{leiner2021sosci} and participants in all conditions were shown the same survey. To ensure that participants had genuinely completed the task and not simply clicked “next”, they were required to upload a screenshot of the recommendation interface before continuing. This step was mandatory, and participants who failed to provide the screenshot were not able to proceed further.
Before measuring each construct in the questionnaire, a short definition of the construct was provided to ensure that all participants shared a common understanding of what is being assessed.
At the end of the survey, participants were asked to respond to several open-ended questions to gather deeper insights into their experiences:
1) Did you have a sense of control when interacting with the recommender system? If so, how?, 2) Did the controllability of the recommender system influence its transparency? If so, how?, 3) Did the controllability of the recommender system influence your trust in it? If so, how?, 4) Did the controllability of the recommender system influence your satisfaction with it? If so, how?, and 5) Would you like to have more control over the system? If so, how?.
These questions allowed us to complement the quantitative measures with qualitative insights, enabling a deeper exploration of participants’ perspectives and expectations regarding different levels of control in the ERS, and how control shaped their perceptions of transparency, trust, and satisfaction.
\subsection{Data Analysis}
Prior to conducting the analysis, the data was cleaned and screened for completeness to ensure that only valid responses were included. 
\subsubsection{Quantitative Data Analysis}
We analyzed the data using Python (pandas, SciPy, and statsmodels libraries). The independent variable (IV) was the level of user control in the ERS, with four between-subjects conditions: \textit{basic, input, process}, and \textit{output contro}l. The dependent variables (DVs) were perceived control, transparency, trust, satisfaction, and perceived quality. For each DV, individual item scores were averaged to create composite scales. The internal consistency of these scales was assessed using Cronbach’s $\alpha$, which indicated acceptable reliability for all measures.
We first computed descriptive statistics (means and standard deviations) for each DV by IV condition and visualized them with boxplots.
Given the between-subjects design with four experimental conditions, we employed one-way analysis of variance (ANOVA) to examine differences in participants’ responses across conditions \cite{st1989analysis}. Where ANOVA was significant, post-hoc tests were conducted to further explore pairwise differences.
Although responses were collected using Likert scales, such data are commonly treated as continuous in behavioral research, allowing the use of parametric tests such as ANOVA and correlation analysis \cite{sullivan2013analyzing}.
To further examine relationships between DVs, we conducted correlation analysis, which provides insights into the strength of the associations between variables \cite{cohen2013applied}.
In addition, we employed structural equation modeling (SEM) to investigate directional dependencies and indirect effects among constructs \cite{kline2016principles}.
\subsubsection{Qualitative Data Analysis}
We analyzed responses to the open-ended questions using a deductive thematic analysis approach, as proposed by \citet{braun2006using}. The analysis was rather deductive as we aimed to find additional explanations to address our research question. In
contrast to inductive (i.e., bottom-up) thematic analysis, which is the data-driven process of coding the data without
trying to fit it into a pre-existing coding frame, deductive (i.e., top-down) thematic analysis is an analyst-driven process
of coding guided by the research questions, allowing the analysis to focus on specific aspects of interest. This form of thematic analysis tends to provide a detailed description of the
data overall and a more detailed analysis of some aspects of the data \cite{braun2006using}. Following this approach, each open-ended question was treated as an initial theme. All responses were first reviewed to gain familiarity with the data and then iteratively coded under the corresponding themes to identify recurring patterns related to participants’ perceptions of control, transparency, trust, and satisfaction.
These recurring patterns were grouped into sub-themes within each condition, capturing common features and attributes (e.g., sorting recommendations, adjusting DNUs) that influenced participants’ perceptions. This process enabled a structured summary of qualitative feedback, highlighting both shared and divergent perspectives across participants.
To ensure reliability, the coding was validated through intracoder agreement following \citet{kuckartz2007einfuhrung}. The same researcher revisited the coding after a time interval, reviewed all assignments, and resolved any inconsistencies.
The identified sub-themes were then used to structure the presentation of qualitative findings, with representative quotes included to illustrate each sub-theme.
\section{Results and Discussion} \label{results}
\begin{figure}[t]
  \centering
  \begin{subfigure}[t]{0.32\textwidth}
    \centering
    \includegraphics[width=\linewidth]{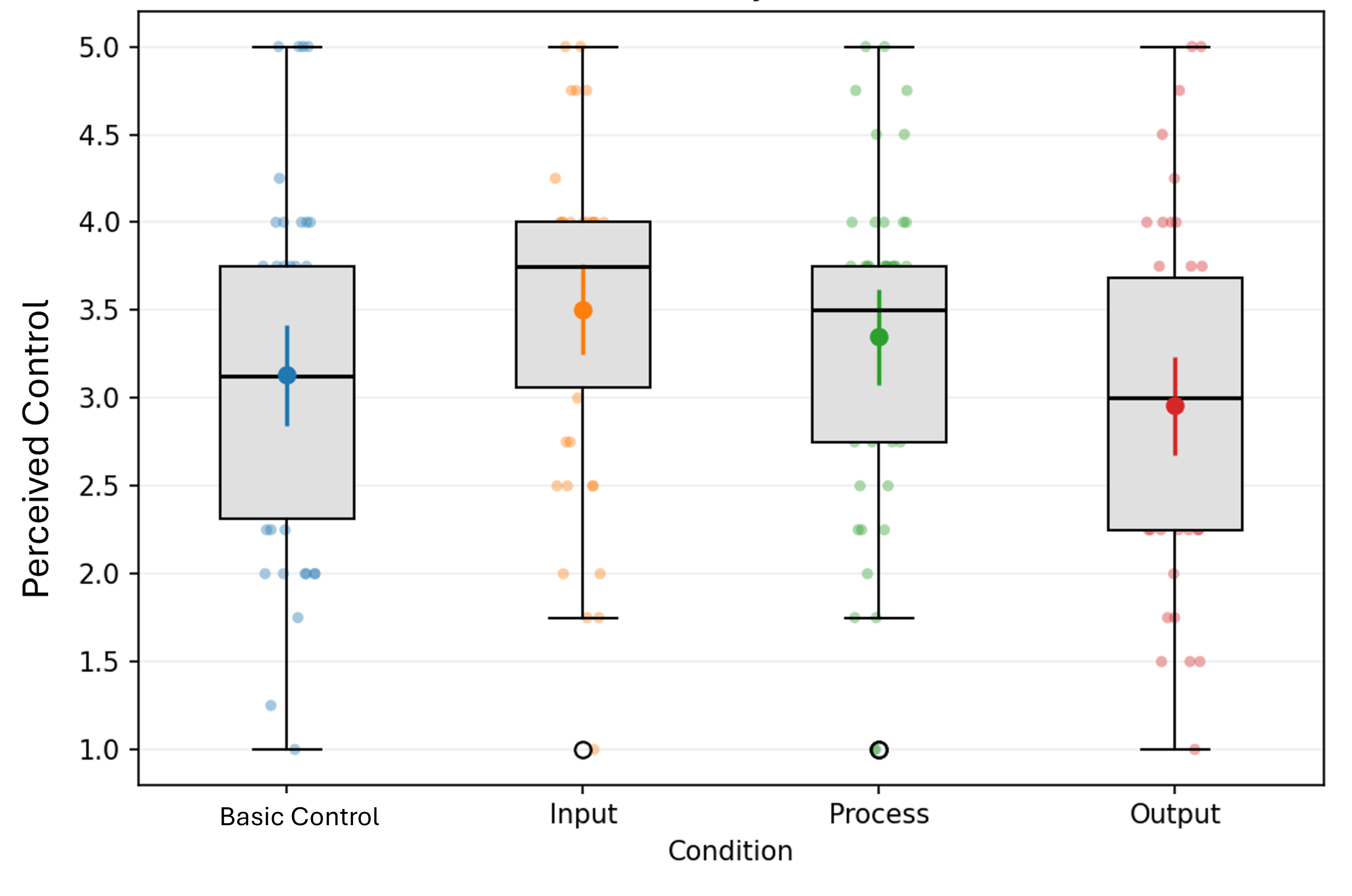}
    \caption{Perceived Control}
    \label{fig:box-control}
  \end{subfigure}\hfill
  \begin{subfigure}[t]{0.32\textwidth}
    \centering
    \includegraphics[width=\linewidth]{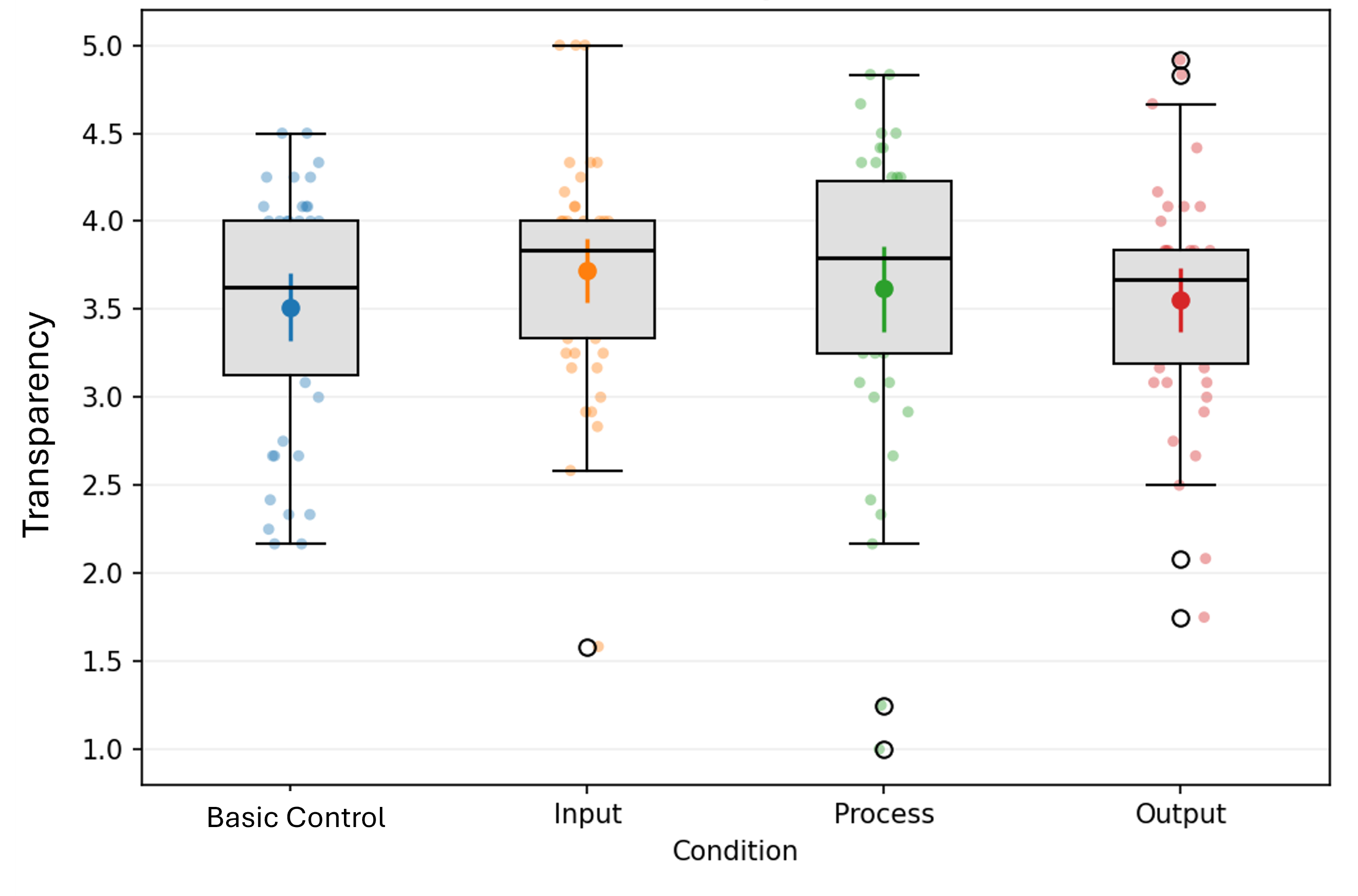}
    \caption{Transparency}
    \label{fig:box-transparency}
  \end{subfigure}\hfill
  \begin{subfigure}[t]{0.32\textwidth}
    \centering
    \includegraphics[width=\linewidth]{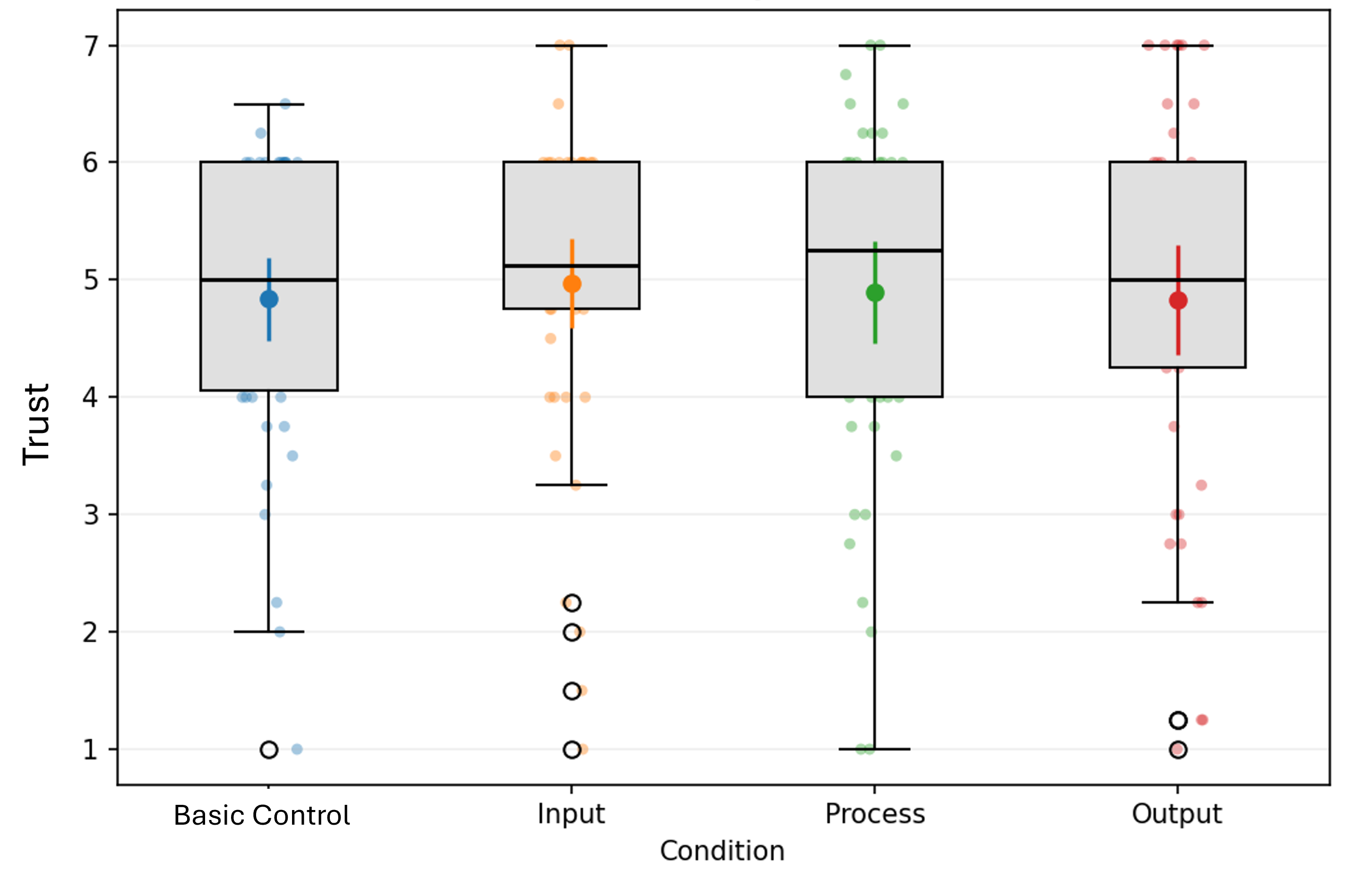}
    \caption{Trust}
    \label{fig:box-trust}
  \end{subfigure}\hfill
  \begin{subfigure}[t]{0.32\textwidth}
    \centering
    \includegraphics[width=\linewidth]{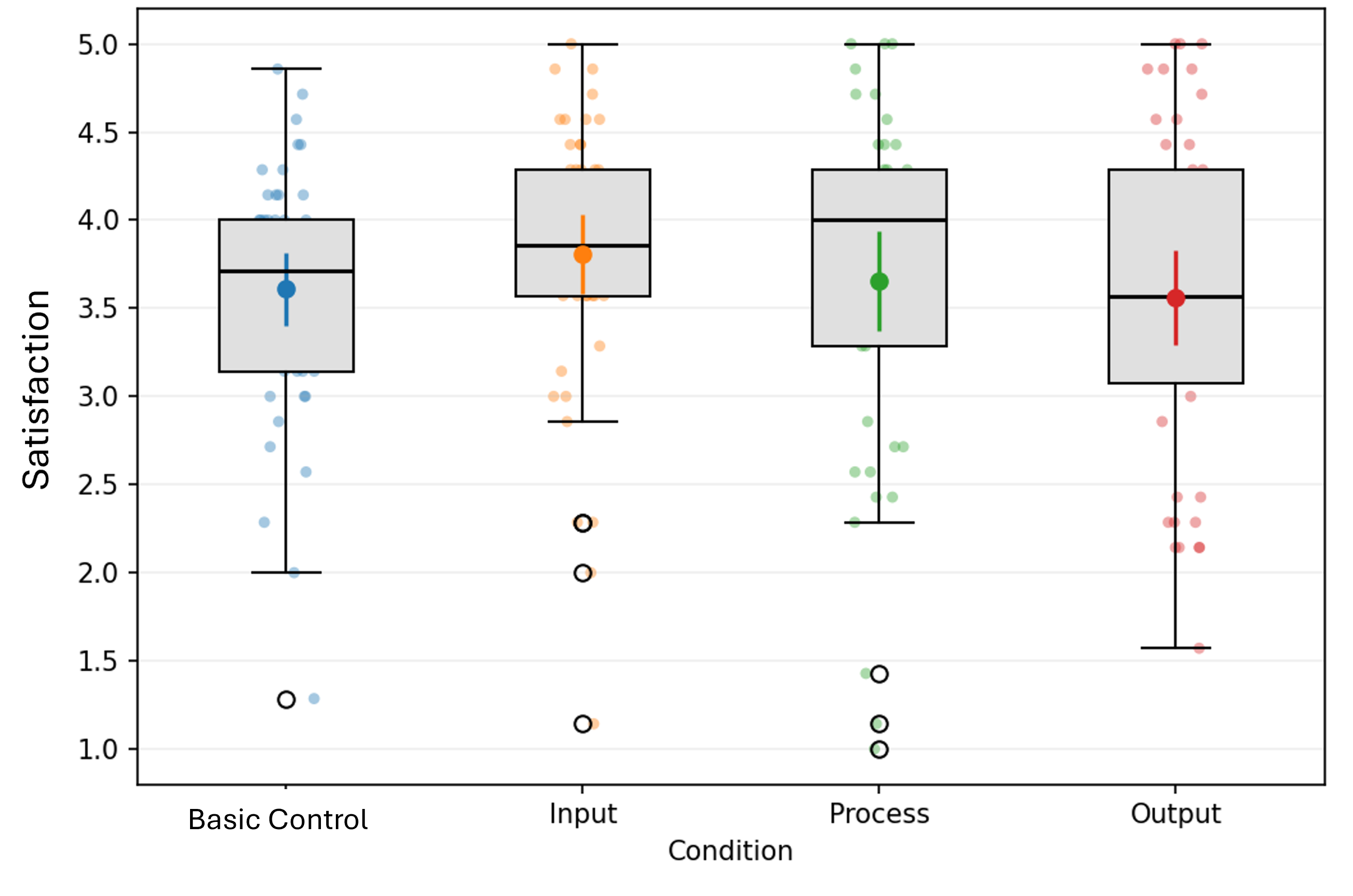}
    \caption{Satisfaction}
    \label{fig:box-satisfaction}
  \end{subfigure}\hfill
  \begin{subfigure}[t]{0.32\textwidth}
    \centering
    \includegraphics[width=\linewidth]{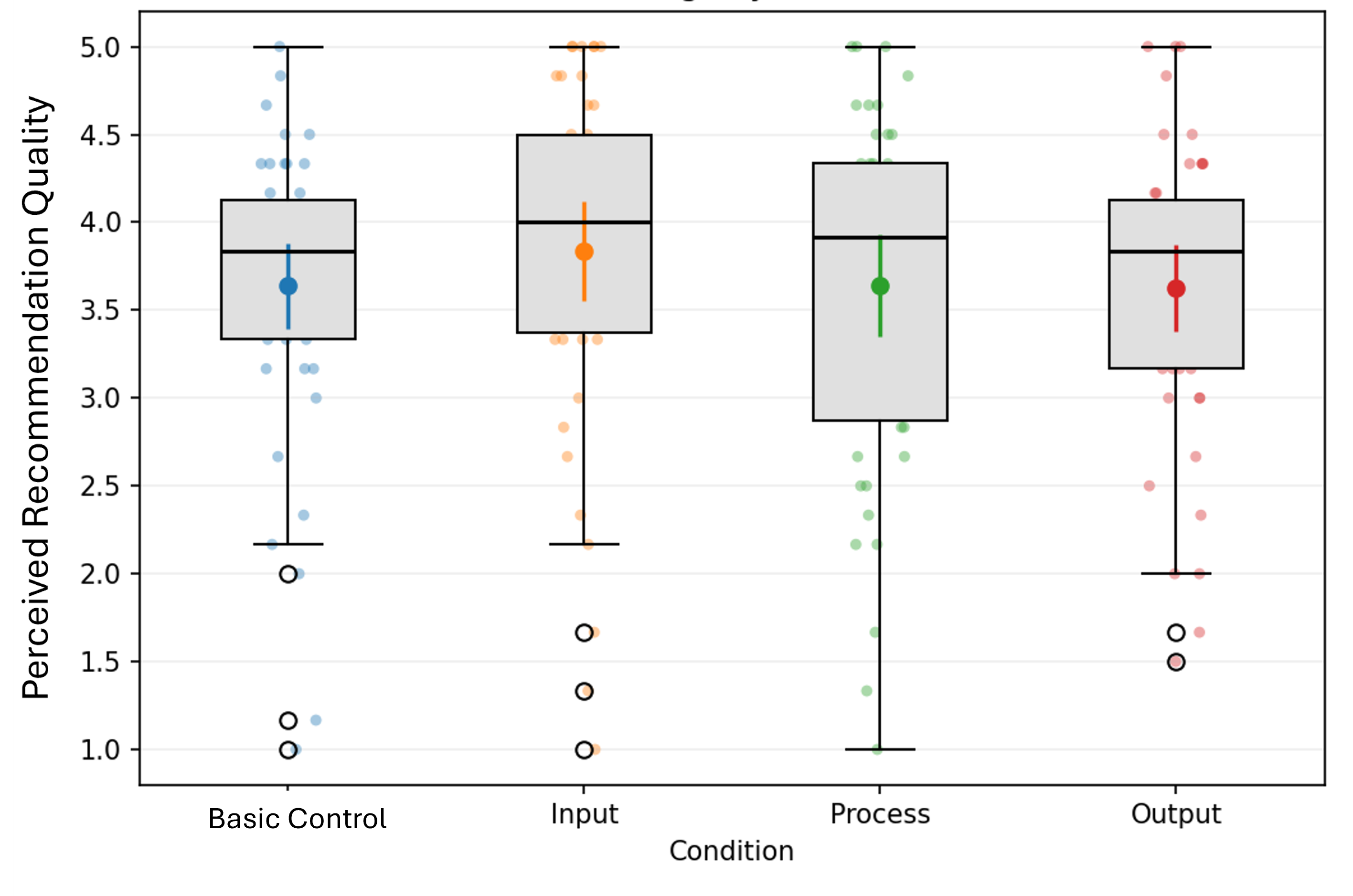}
    \caption{Perceived Recommendation Quality}
    \label{fig:box-quality}
  \end{subfigure}\hfill
  \caption{Boxplots of all dependent variables (DVs) by condition (\textit{basic control, input, process, output}). 
  Each plot shows the median (box center line), interquartile range (box), whiskers, 
  jittered individual responses, and group means with 95\% confidence intervals (CIs).}
  \label{fig:all-boxplots}
\end{figure}

We first examined the overall distribution of responses across different levels of control (conditions). Table \ref{tab:descriptives} reports the descriptive statistics (means and standard deviations) for each DV (i.e., perceived control, transparency, trust, satisfaction, and perceived recommendation quality) per condition. Figure \ref{fig:all-boxplots} illustrates the distributions with boxplots. 
Across all DVs, the group means were relatively close, with only small variations between conditions. For perceived control, scores were highest in the \textit{input control} condition (M = 3.51, SD = 0.89) and lowest in the \textit{output control} condition (M = 2.96, SD = 0.97), while the \textit{basic control} and \textit{process control} conditions fell in between. Transparency ratings were similar across groups, ranging from 3.51 (\textit{basic control}) to 3.74 (\textit{input}), suggesting little difference in perceived transparency. Trust scores were also moderate to high in all conditions, with means between 4.72 and 4.97, and showed minimal variation. Satisfaction was also comparable across groups, with means between 3.56 (\textit{output}) and 3.81 (\textit{input}). Finally, perceived recommendation quality displayed very similar averages across conditions, ranging from 3.63 to 3.83. Overall, the highest value of mean for each DV in all conditions is the \textit{input control}. The standard deviations indicate a moderate level of variability in participants’ responses within each condition, with values generally ranging between 0.7 and 1.1. Trust showed relatively larger variability (SDs up to 1.50), suggesting greater diversity in how participants evaluated the system’s trustworthiness. 
These descriptive findings provide a general overview of participants’ responses and serve as a basis for the inferential analyses reported in the following sections related to research question.
\begin{table}[t]
\centering
\caption{Descriptive statistics (Mean $\pm$ SD) for each dependent variable (DV) by condition (N = 46 per condition).}
\label{tab:descriptives}
\begin{tabular}{lcccc}
\toprule
Dependent Variable (DV) & Basic Control & Input Control & Process Control & Output Control\\
\midrule
Perceived Control              & 3.13 $\pm$ 0.98 & \textbf{3.51} $\pm$ 0.89 & 3.35 $\pm$ 0.95 & 2.96 $\pm$ 0.97 \\
Transparency                   & 3.51 $\pm$ 0.93 & \textbf{3.74} $\pm$ 1.10 & 3.56 $\pm$ 1.05 & 3.68 $\pm$ 1.02 \\
Trust                          & 4.84 $\pm$ 1.21 & \textbf{4.97} $\pm$ 1.32 & 4.89 $\pm$ 1.50 & 4.72 $\pm$ 1.27 \\
Satisfaction                   & 3.61 $\pm$ 0.71 & \textbf{3.81} $\pm$ 0.79 & 3.65 $\pm$ 0.98 & 3.56 $\pm$ 0.92 \\
Perceived Recommendation Quality & 3.64 $\pm$ 0.84 & \textbf{3.83} $\pm$ 0.98 & 3.64 $\pm$ 1.00 & 3.63 $\pm$ 0.85 \\
\bottomrule
\end{tabular}
\end{table}
\subsection{Differences Among Levels of Control}
To answer RQ1: "How do different levels of user control in an ERS affect users’ perceived control, transparency, trust, satisfaction, and perceived quality of recommendations?", we conducted one-way ANOVAs with the four conditions (\textit{basic control, input, process, output}) as the independent variable and each dependent variable (perceived control, transparency, trust, satisfaction and perceived quality) as the outcome. The results are presented in Table \ref{tab:rq2_anovas}. Homogeneity of variances was verified with Levene’s test. Where ANOVAs were significant, we followed up with Tukey’s HSD post-hoc tests.
\begin{table}[t]
\centering
\caption{ANOVAs across conditions (\textit{basic control, input, process, output}). Reported are $F-statistics$, $p-values$, and effect sizes ($\eta^2$).}
\label{tab:rq2_anovas}
\begin{tabular}{lrrr}
\toprule
DV & $F$ & $p$ & $\eta^2$ \\
\midrule
Perceived Control & 2.980 & \textbf{0.033} & 0.047 \\
Transparency      & 0.759 & 0.518 & 0.012 \\
Trust             & 0.090 & 0.965 & 0.002 \\
Satisfaction      & 0.711 & 0.547 & 0.012 \\
Perceived Quality & 0.535 & 0.659 & 0.009 \\
\bottomrule
\end{tabular}
\end{table}

\begin{table}[t]
\centering
\caption{Tukey's HSD post-hoc comparisons for Perceived Control. Reported are mean differences, adjusted $p$-values, and significance. Sig: * = significant, ns = not significant.}
\label{tab:rq2_tukey_pc}
\begin{tabular}{llrrc}
\toprule
Group 1 & Group 2 & Mean diff & $p_{adj}$ & Sig \\
\midrule
Input & Basic control    & -0.375 & 0.232 & ns \\
Input & Output  & -0.549 & 0.030 & * \\
Input & Process & -0.158 & 0.855 & ns \\
Basic control  & Output  & -0.174 & 0.815 & ns \\
Basic control & Process &  0.217 & 0.690 & ns \\
Output & Process & 0.391 & 0.199 & ns \\
\bottomrule
\end{tabular}
\end{table}

\subsubsection{Effects on Perceived Control}
The ANOVA for perceived control was significant (p = .033, $\eta^2 = .047$). This means that the different levels of user control had a significant effect on users' perceived control. 
To further investigate which particular level of control influenced perceived control, we conducted Tukey's HSD test which performs all possible pairwise comparisons to see which specific groups are different from each other, as presented in Table \ref{tab:rq2_tukey_pc}.
The comparison between the Input and Output control groups showed a statistically significant difference ($p_{adj}$=0.0304). This means that the level of perceived control was significantly different for users in the Input group compared to those in the Output group.
The descriptive statistics (Table \ref{tab:descriptives}) and the boxplot (Figure \ref{fig:box-control}) confirms this result, with Input scoring the highest and Output the lowest. All other pairwise comparisons were not statistically significant ($p_{adj}$>0.05). 

Qualitative responses further contextualized these findings. In response to the question “Did you have a sense of control when interacting with the recommender system? If so, how?”, most of the participants in the \textit{basic control} condition (N=35, 76\%) frequently reported that they felt in control  because, "\textit{the recommendations were based on the concepts they marked as not understood themselves}" (N=23), but did not feel any control beyond that. For instance, P17 reported: "\textit{The recommendations were based on my not understood concepts, however, I did not have much control over how the system selected the videos, or the order in which they were shown}", and P21 mentioned that, "\textit{I was being able to select which concepts I'm unsure of, but beyond that I don't have much control over what is recommended to me}". These results suggest that participants already held positive perceptions of the ERS without the additional controls of the ERS' input, process, or output. One possible explanation is that the system provided participants with the possibility to mark concepts as not understood to generate the recommendations, which may have fostered a sense of control of the ERS.

Nearly all participants in the \textit{input control} condition (N=45, 98\%) reported experiencing a sense of control. A central reason was their ability to directly manipulate the system’s input through the provided options such as adjusting concept weights with sliders and selecting which concepts should be used to generate the recommendations. For example, P51 explained, \textit{“Yes I did. Due to the fact that I could change the weightings of the concepts I was trying to understand, it meant video recommendations would be altered to provide a better mix. It allows content to be learned with priorities”}. Similarly, P88 described experimenting with the sliders and concepts as, \textit{“Yes, I played with the sliders to see what differences it made and with different concepts to see how it changed the recommended videos. I did this a few times with different concepts”}, 
and P84 emphasized the flexibility of the input options and how they impact improving the recommendations, \textit{“I was able to choose how I wanted the recommendations to be, if it was for the current slide, all slides, or manually and how they provided better and more accurate videos”}. 
When further asked whether they desired even more control, most participants responded that the system already provided sufficient options. Where additional control was suggested, it mainly concerned filtering the videos or providing feedback to them, rather than altering the input control itself. As P55 reflected, “\textit{Overall, the way the concepts could be selected as input to create recommendations gave a very high level of control. I am unsure how greater control could be implemented}”. Together, these findings illustrate that participants felt in control primarily because they could actively control the system’s input. 

In the \textit{process control} condition, most participants (N = 38, 82\%) reported experiencing a sense of control. They described feeling in control because they could influence how the recommendations were generated and adjust sliders to manipulate the ranking of recommended videos. For example, P101 noted, “\textit{Yes, I noticed that the video recommendations were changing according to how I interacted with the slider to manipulate ranking, rather than showing the same videos for different inputs}”. Similarly, P105 emphasized the benefits of fine-tuning as, “\textit{Yes, I was able to fine-tune how I wanted the recommended videos to be generated and was able to find much more relevant videos after changing that}”. P138 highlighted the ability to customize ranking criteria as, “\textit{I enjoyed playing around with the scales of factors that impact the recommendations since I could rank by various options such as views on YouTube and likes to filter out unwanted content. Those were really good to see the differences of recommended videos}”. 
When asked whether they desired even more control, participants were split. About half (N = 23) stated that the current level of control was sufficient, noting that additional options might complicate the experience. For instance, P105 mentioned, “\textit{No, I think the current amount of control is more than enough. Maybe more control would be too much}”). The other half expressed interest in further control features, particularly filtering unwanted videos based on specific criteria (e.g., YouTube channels, video duration, or genre) and the ability to provide feedback on recommendations (e.g., liking or disliking videos).

In the \textit{output control} condition, most participants (N = 33, 72\%) reported experiencing a sense of control. They described this control as stemming from their ability to influence how recommendations were presented, for instance by sorting videos or providing feedback. As P182 explained, “\textit{I could sort the videos based on various criteria like views, likes, etc., which helped}”. Similarly, P175 noted, “\textit{Yes, as I could say what was helpful and what was not}”.
Some participants also mentioned that being able to save videos for later contributed to a sense of control, e.g., P158 mentioned, “\textit{I could decide which videos to watch now or later to help me understand the concepts better}”. However, some participants linked their perceived control to the recommendations being based on the concepts they did not understand when they started using the system, as reflected by P156, “\textit{The videos given to me were based on what I didn’t understand so I felt in control}”.
When asked whether they desired additional control, more than half of the participants (N = 27) stated that the current level was sufficient and that more options were unnecessary. For example, P144 remarked, “\textit{I do not feel like I need more control. I feel like giving it the information I need and then allowing it to do what it needs to do to get me the resources I need is enough for me}”. Similarly, P175 commented, “\textit{No, I was quite happy with the amount of control. If you had more, it may complicate things more}”. Among those who did wish for more, suggestions focused on removing unwanted videos, controlling the language of results, or providing more granular feedback. For example, P178 suggested, “\textit{I think having an option to remove certain results would be beneficial, so I can remove all videos about that song to further narrow down the best videos for me}”. Likewise,
P168 emphasized the need for finer-grained feedback and stated, “\textit{While the current controls like sorting and saving videos are helpful, I’d like to be able to provide more granular feedback. I would prefer a way to explicitly dislike a video from a specific channel, so the system understands what content I find irrelevant or unhelpful}”.

Answering the question, "would you like to have more control over the system? If so, how?", most of the participants (N=33, 72\%) in the \textit{basic control} condition expressed a desire to have more control over the ERS. Most of the desirable control options were related to the ERS input (e.g., search, add, or remove concepts to be used for recommendation). 
Moreover, many participants (N=21, 45\%) in the \textit{output control }condition desired to have more control over the ERS. Most of them (N=13) desired control over the ERS input (e.g., \textit{"more ability to pick and choose concepts to search"} (P142), \textit{"searching for topics by keyword"} (P151)). This can explain the results that the Input vs. Output pair was significantly different in terms of perceived control (see Table \ref{tab:rq2_tukey_pc}).   

Overall, the ANOVA results revealed a significant effect of level of control on perceived control. Moreover, the quantitative and qualitative results indicate that, while process and output controls also offered meaningful interaction, participants perceived the most control when they could directly influence the system’s input. The post-hoc tests further showed that input control provided a higher sense of controllability of the ERS compared to output control. 
A likely reason is that the user profile lays the foundation of the recommendation process. 
By selecting and weighting their DNU concepts, participants could actively create and edit their own user profile used as input to generate recommendations. This direct interaction with the user profile not only fostered a stronger sense of control but also reinforced the perception that the system genuinely reflected their learning needs.
These findings suggest that future ERS designs should prioritize input level control to help users control their user profile, thus increasing their perceived control of the ERS.  
\subsubsection{Effects on Transparency}
Transparency ratings did not significantly differ between conditions in ANOVA (p = .518, $\eta^2$ = .012). The mean scores were relatively high across all groups, suggesting that participants generally considered the system transparent regardless of the specific control mechanism.
The qualitative data further explained how participants interpreted transparency. When asked “Did the controllability of the recommender system influence its transparency? If so, how?”, (N=29, 63\%) of participants in the \textit{basic control} condition agreed to it. The reason mentioned by most of the participants was that the system informed them what the recommendations were based on, for instance, P2 mentioned that, \textit{"on the recommended videos page it states that the recommendations I got are based on the topics I did not understand which makes the recommender system a bit more transparent".} Similarly, P32 attributed transparency to control by saying that, "\textit{if I have more control it would feel more transparent}". 

Furthermore, most participants in the \textit{input control} condition (N = 39, 85\%) reported that the Input control features enhanced their sense of transparency. Many equated transparency with being able to see and understand why recommendations were generated, since they were explicitly tied to the learner’s selected DNU concepts. As P51 explained, “\textit{The amount of controllability made me always super aware of why certain videos are being shown. It is purely based on concepts you are trying to understand and how much you want each concept to be weighted in the video recommendation}”, and P72 remarked that, \textit{"Because I can control some aspects it makes it easier to understand
why it is making the recommendations"}.
Similarly, P66 remarked, “\textit{I feel that it does influence its transparency because the level of control gives you ownership over the results and that ownership makes you understand what is going on}”. In the same vein, P53 stated, “\textit{You’re able to adjust things to your liking … which are displayed with the recommended videos on the left side so how the system works is apparently shown to the end user}”. This finding is in line with previous works showing a positive relationship between controllabity of the RS input and transparency. This line of work distinguishes transparency with respect to input (i.e., what data does the RS use?) and stresses the importance of enabling transparency by opening and controlling the typically black box user profiles, that serve as the RS input. For example, \citet{jin2016go} argues that transparency is supported by letting the users view and interact with data of the user profile used in different steps of the recommendation process. In the same vein, \citet{Donovan} noted that users feel more informed when they can control and visualize how their inputs affect the system’s output, leading to higher perceptions of transparency \cite{he2016interactive}. In fact, providing user control on the RS input can support users in better understanding an important part of the underlying mechanisms of the RS, namely the user profile data used to make recommendations. It hence contributes to explaining the inner working of the RS, and thus increases transparency \cite{graus2018let}. 

In the \textit{process control} condition, most of the participants (N = 35, 76\%) reported that controllability enhanced their perception of transparency in the system. Many highlighted the ranking mechanism via sliders as the primary feature that made them feel the system is transparent, since adjusting multiple factors and immediately observing changes in recommendations gave them insight into how the system works. As P109 explained, “\textit{Yes, it was very transparent, as it showed me various factors, likes by other users, etc., gave me the ability to pick what I wanted by moving the slider on the side to best fit my tastes, and I noticed that the video recommendations were changing according to how I interacted with the slider}”. Some participants found switching between algorithms as helping them to understand its working, as P127 remarked, “\textit{Yes, it's transparent because you can control how the videos are generated}”. P98 stated, “\textit{I think it allows you to tailor recommendations, so it is transparent in that sense, as you are deciding the parameters of the results}”. Some of the participants in this condition found the system to be transparent just because they knew what the recommendations are based on, for instance P120 reported, "\textit{I can understand the recommendations are based on the concepts I am interested in, I can understand why they are chosen}". These results show that controlling the process can help users understand how the ERS works. 
This is in line
with other studies in the IntRS domain showing that providing insight into the recommendation process through controllability can lead to increased transparency \cite{tsai2017providing, tsai2021effects}.

In the \textit{output control} condition, many participants (N = 30, 65\%) also felt that controllability increased transparency. Some attributed this to the fact of having control, linking it to an impression of openness. For instance, P140 stated, “\textit{Yes. The more control people have access to, the more transparent they find the system. They will feel like there’s nothing to hide because of the degree of control}”). Others referred to specific features such as providing feedback on videos (e.g., P175 mentioned, “\textit{By saying if it’s helpful or not it then helps the system to find better options}”), viewing similarity scores (P158 stated, “\textit{It showed the relevance of each video to the concept I was searching for}”), or sorting and filtering options (e.g., P182 said, “\textit{it was transparent in how it ordered the videos}”). Few participants, however, grounded their sense of transparency simply in the fact that recommendations were tied to their DNU concepts, regardless of the additional output controls. For example, P163 explained, “\textit{It makes it clear that I can get the recommendations I want by choosing certain concepts}”.

Overall, our results show that participants considered the
system transparent at different levels of control. This suggests that it is necessary to view transparency as a multi-faceted concept. As pointed out by \citet{hellmann2022development}, transparency can be
distinguished with respect to input (what data does the system use?), process (how and why is an item recommended?),
output (why and how well does an item fit one’s preferences?). Our results further show that transparency, which is often linked to users’ understanding of a system’s inner logic and supports users in building an accurate mental model of how the system works \citep{guesmi2023justification, ngo2020exploring}, is not only related to controlling the process (i.e., how the system works), but can also relate to users' control of the RS input (i.e., user profile) or the output (i.e., recommendations). This indicates that user-perceived transparency can also be high, when users are allowed to control their input profile or the recommendations.       
Our findings suggest that it is important to differentiate between objective and user-perceived transparency, as also stressed in the literature on explainable recommendation \cite{gedikli2014should,guesmi2023justification}. In the IntRS domain, on the one hand, objective transparency means that the IntRS reveals all the information about its inner logic and allows users to interact with the recommendation process. On the other hand, user-perceived transparency is based on the users’ subjective opinion about the extent to which they perceive such information is available and feel that they understand the meaning of the provided information.
Therefore, we suggest that it is essential for an ERS to provide transparency by exposing and interacting with the user profile data behind a recommendation as well as allowing users to manipulate the recommendation process and output. 
Noticeably, our qualitative results showed that the majority of participants found control with the input and process to be more influential to the transparency of the ERS. Thus, we can conclude that transparency is more related to the input or process, rather than the output of the ERS.
Additionally, a large number of participants across all control conditions attributed transparency to the fact that they knew what the recommendations are based on, namely their DNU concepts. This indicates that letting users control and scrutinize their user profile is sufficient to improve users' perceived transparency.  

\subsubsection{Effects on Trust} \label{control_trust}
The ANOVA for trust did not yield significant differences across conditions (p = .965, $\eta^2$ = .002). However, trust ratings remained consistently moderate to high regardless of control level.
Our qualitative results provide insight into this stability. Regarding whether controllability influences their trust in the ERS, 55\% (N=26) of the participants in the \textit{basic control} condition agreed to it. 
Many participants directly linked trust to control mentioning that, "\textit{yes, I trust it as I can control it}" (P1), "\textit{Less control means that I have less trust in it, as it does not allow me to give it more information to get the best results}" (P26), and "\textit{I would trust it more if I were able to control more}" (P27). 
10 participants trusted the system because they think that the recommended videos are relevant as they are based on the concepts they did not understand so they know the reason behind the recommendation. For instance, P33 remarked: "\textit{I like the idea of being able to select topics and then have a system recommend me things specific to what I don't understand}". P17 mentioned that, "\textit{Being able to mark concepts I didn’t understand made me trust that the recommendations were relevant to my learning}". Similarly, P24 stated, "\textit{Yes I feel like if I consistently get relevant results based on the concepts I did not understand, I can trust the system}". 

In the \textit{input control} condition, most participants (N = 34, 74\%) reported that controllability positively influenced their trust. Many 
highlighted the ability to fine-tune their DNU concepts, which increased their confidence in the system as, “\textit{It makes me trust it more as I could select/unselect concepts and edit each concept’s importance to be used for recommendation to get desired results}” (P52). Similarly, P65 mentioned, "\textit{I know how the recommendations are generated when I can change my concepts, that made me trust it}". The participants further attributed trust to the fact that their control of the ERS input enables them to become aware of system errors and consequently helps them give feedback and correction in order to improve future recommendations. For instance, P76 mentioned, "\textit{I know that if the first videos recommended are not exactly what I want then I can edit certain concepts to get some more focused videos}". Our results show that allowing users to control their profile in an ERS has several benefits related to trust. Interacting with the ERS input (i.e., user profile) allows users to
understand what data does the ERS use for providing recommendation, and why and how well does a recommendation fit their user profile. In addition, users can detect wrong assumptions made by the system, correct their profiles when they disagree with (parts of) it in order to adjust the recommendation results according to their needs and preferences. The positive influence of input control on trust aligns with findings by \citet{ooge23steering} that allowing students to adjust their mastery level to recommend exercises and to visualize its impact was positively perceived to build their trust in the ERS. 
This is further inline with the findings from \citet{schaffer2015hypothetical} that profile update tasks  improve perceived trust in the recommender, regardless of actual recommendation accuracy. In a similar vein, \citet{loepp2014choice} found that users trust the system when they know that it takes only their needs into account. Moreover, \citet{bruns2015should} demonstrated that control over input can enhance trust.

In the \textit{process control} condition, many participants (N = 30, 65\%) also reported that controllability enhanced their trust in the ERS. Some participants attributed their perceived trust to the fact that the algorithm is transparent to them. For example, P101 mentioned that, "\textit{It makes me feel I am controlling my own recommendations rather than being entirely reliant on a behind-the-scenes algorithm, this enhances trust}". 
The participants also valued the options provided to control the process. For instance, P126 stated, “\textit{I trusted it more because it gave me more options to control how the recommendations are generated}”.  
Similarly, P98 mentioned, "\textit{I guess it improves my trust a bit because of the amount of options and being able to be in control of what boxes are checked, how much weight they have, etc. It helps improve trust because it felt like everything was being considered and shown to me, at least on my first impression}".  
In contrast, some participants found controlling the process too much, as mentioned by P120, "\textit{The platform is a bit intimidating}".  
These results suggest that different users have different needs for control, as also pointed out by \citet{jin2018effects}. Too much control options can significantly influence cognitive load \citep{jin2018effects} which often results in decreasing trust \cite{samson2015effects, ahmad2019trust}. 
This further highlights that there is a trade-off between the amount of user control and the level of perceived trust users develop when interacting with the recommendation process.   
Our findings are in line with results of previous research on explainable recommendation and XAI showing that the provision of additional explanations increases cognitive effort which can result in decreasing trust \cite{zhao2019users, chatti2022more, kulesza2015principles, kulesza2013explanations, yang2020visual, kizilcec2016much}.
This line of research concludes that designing for trust requires balanced system transparency: “not too
little and not too much” \cite{kizilcec2016much} and “be sound”, “be complete” but “don’t overwhelm” \cite{kulesza2013explanations, kulesza2015principles}. 

In the \textit{output control} condition, most participants (N = 34, 74\%) reported that control features influenced their trust. Output controls such as sorting, saving, and feedback were commonly cited as fostering confidence in the recommendations. As P182 explained, “\textit{Using the sort feature helped me to get better results and this increases trust}”. Likewise, P168 valued the ability to save content for later as, “\textit{Yes, I trust it … the fact I can save videos for later is a big help}”. In the similar vein, P158 mentioned, "\textit{I felt like I had enough control to be able to trust the system and what it recommended to me. It also allowed me to choose which videos to watch later and rate them on helpfulness}".  
The majority of the participants trusted the system when it provided relevant recommendations that meet their needs, as P153 remarked, “\textit{I trust the recommender system as long as it’s providing me relevant recommendations from credible sources, which I believe it is doing}”. Similarly P179 said, "\textit{I trust it because it provides me relevant videos that help in learning}".

Overall, our study results show that perceived trust in the ERS is a multidimensional concept which can be influenced by different factors, including awareness of the data used for recommendation, users’ ability to inspect and modify their user profile, controlling the recommendation process, and usefulness of the recommendation results. Our qualitative findings further show that trust is positively influenced by the presence of control mechanisms at different control levels (i.e., input, process, and output), but was most strongly rooted in input and output controls. This suggests that it is beneficial to empower users to control their user profiles and provide relevant recommendations as a means to enhance trust in the ERS. Moreover, designing for trust requires a careful balance, learners should be empowered to influence the recommendation process, but without overwhelming them with too many control options.   

\subsubsection{Effects on Satisfaction}
The ANOVA showed no significant differences in satisfaction across control levels (p = .547, $\eta^2$ = .012). In our descriptive analysis, the mean values indicated high satisfaction across all conditions, indicating participants’ satisfaction with the ERS at different levels of control.
Our qualitative results reinforced these findings. Most of the participants in the \textit{basic control} condition (N=33, 72\%) reported that they were satisfied with the ERS because they were able to control it. One of the major reason reported by most of the participants was that because they knew what the recommendations are based on. For instance, P34 stated, "\textit{it being based on my needs means the videos are relevant and satisfy what I was looking for}", and P36 mentioned, "\textit{because I was able to see several different options of videos based on what I didn't understand}". 

In the \textit{input control} condition, the majority of participants (N = 41, 89\%) expressed satisfaction with the ERS, often linking it to the provided control options and their impact on improving the recommendation results. For instance, P54 mentioned, "\textit{I am satisfied yes. I think there is enough flexibility and scope for user-input, to search a wide range of videos. The system does it well}". P69 reported, "\textit{Yes because I was getting different results based on how I changed the values of concepts I didn't understand}". Similarly, P80 noted, "\textit{being able to go back and change exactly what is being searched for on the fly is very helpful}" (P80). Related to their satisfaction based on the quality of the recommendation results, P66 reported, “\textit{The controllability influences my satisfaction positively because it stop me from seeing unwanted content}”. And, P89 said, “\textit{it showed good recommendations}”. 
These results confirm findings from earlier studies showing that exploring users' own profile interactively increases their satisfaction with the RS \cite{linkedvis2013}. 
Users could also relate their satisfaction to their user experience as P48 elaborated, "\textit{I am satisfied and I enjoyed using the platform as a whole and I can see how it would be useful for a user to learn from}". In the similar vein, participants showed their willingness to use it in future, as P92 mentioned, "\textit{I was satisfied with the system and would consider using it again}". This finding suggests that the system effectively influenced users' decisions to continue using the ERS and reflects their satisfaction with the ERS as according to \citet{tintarev2007}, satisfaction can also be measured indirectly, measuring user loyalty. Thus, users’ use intentions can be seen as an indirect measure of satisfaction with the system. 

In the \textit{process control} condition, most participants (N = 37, 80\%) reported satisfaction, largely due to being able to fine-tune the system’s behavior. For instance, P114 mentioned, “\textit{It does influence my satisfaction by letting me use my personal preferences. If my preferences are used then I know I am more likely to be satisfied}”. Similarly, P125 explained, “\textit{I am satisfied because if it is off from my needs I can fix it myself}”, while P100 elaborated, "\textit{It made me more satisfied because it felt like the system could understand my needs}".
Many participants emphasized that simply being able to manipulate the process helped them find more relevant videos, which further added to satisfaction, as P99 explained, “\textit{Yes, practically the more you can control, the better the results and output is}”. Similarly, P96 stated, “\textit{Yes, as the videos are made more relevant to me}”. This controllability was even described as a decisive factor for some as P118 stated, “\textit{I’m satisfied because I can control. If I had no control over what was recommended, I wouldn’t use it}”. Particularly, participants valued the ability to control ranking and algorithms, as P138 remarked, “\textit{I’m very satisfied with it this so far … being able to change the algorithm for what I want to learn is more than enough for me. I appreciate the power I gain from it}”. These results are inline with the findings from other studies highlighting that interacting with the process of the RS is also found to be influential to user satisfaction \cite{tsai2017providing,tsai2021effects}\cite{jung2019video}. 

Similarly, providing control over the \textit{output} of the ERS lead to an increased satisfaction (N=39, 85\%). 
Participants highlighted the ability to provide feedback, sort, and save videos as features that directly improved their satisfaction. For instance, P126 commented, “\textit{I was more satisfied having control options, and being able to rank it helpful and not helpful as it felt like I was able to filter out useless content}”. Similarly, P139 remarked, “\textit{Yes, as I can adjust it until I find videos I like that can help me, and downgrade videos that aren’t helpful}”. Specific features such as sorting and saving were praised too as P168 commented, “\textit{The ability to sort videos … and save them for later gave me a sense of control over my learning process. This personalized interaction made the system feel more tailored to my needs, which directly increased my overall satisfaction}”. Similarly, P171 added, “\textit{I like how I can say if the video was helpful or not}”. Participants emphasized the value of output control for shaping their recommendations, as P143 mentioned, “\textit{Yes, since I had control over the types of videos I want to and I can sort them to get desired result, this makes me satisfied}”. In the same vein, P168 highlighted, "\textit{yes, it seems to give me better videos with the filter}". 
Ease of use was also a positive factor, as P147 noted, “\textit{Yes it does because it is easy to use which makes me satisfied}”. P164 similarly explained, “\textit{It was simple to use so anyone could figure it out and the control we had was okay and made me feel satisfied}”. 
These results are inline with the findings from previous studies showing that providing users with the ability to control and interact with the recommendations positively impacts their satisfaction with the RS \cite{smallworlds, wong2011diversity,talkexplorer, 2012tasteweights,knijnenburg2012inspectability} and that user control over the output generates more accurate recommendations that match users' interests \citep{maxwell2015}. 

Overall, satisfaction was high across all four conditions, consistently tied to the the ability to exercise control over the ERS, awareness of what the recommendations are based on, relevance and usefulness of the recommendations, and ease of use.   
\subsubsection{Effects on Perceived Quality}
The ANOVA did not reveal significant differences across conditions, (p = .659, $\eta^2$ = .009), suggesting that perceived quality was rated similarly across groups. The descriptive statistics, however, indicated that the \textit{input control} condition yielded the highest mean ratings, with the other conditions showing comparable values.
Our qualitative results provide further insights into these results.
Most participants in the \textit{basic control} condition reported that the ability to mark DNU concepts helped them obtain relevant and useful recommendations, which positively influenced their perception of recommendation quality. For instance, P1 stated, “\textit{I could tell it what I need help with, and it found the relevant videos}”.
Similarly, P33 highlighted that “\textit{the recommended videos were directly related to the topics I chose},” and P15 mentioned, “\textit{I could select which concepts I did not understand and it gave me videos that helped me understand the material better}”. 

In the \textit{input control} condition, many participants highlighted that customization options improved the quality of recommendations by reducing irrelevant results and tailoring content to their learning needs. As P74 explained, “\textit{The more customization options there were, meant there were less irrelevant video options I had to search through}”. Similarly, P51 elaborated, “\textit{Due to the fact that I could change the weightings of the concepts I was trying to understand, it meant video recommendations would be altered to provide a better mix. It allows content to be learned with priorities}”. Participants also emphasized personalization as a driver of quality, with P57 stating, “\textit{it felt more personalized and useful}”. 
Overall, participants valued how input control enabled them to tune the system to their preferences and quickly find relevant resources. These results confirm earlier findings from \citet{knijnenburg2012explaining} suggesting that allowing users to explicitly control input data leads to increased perceived quality. Similarly,
\citet{jin2016go} \cite{jin2017different} found that controlling the input profile or modifying the weights of data being used for recommendations leads to better perceived recommendation quality.

In the \textit{process control} condition, participants also reported that controllability improved recommendation quality by allowing them to fine-tune the recommendation results. For example, P133 noted, “\textit{Yes, because you are able to tailor the recommendations to your own tastes and goals, leading to more relevant or interesting suggestions}”. Similarly, P105 highlighted the effect of sliders and stated that, “\textit{Having the ability to fine tune the recommended videos allowed me to find much more relevant videos after making some changes in the recommender system}”. Other participants emphasized the link between personalization and recommendation quality. For example, P121 stated “\textit{Yes, I know I will get videos more tailored to my learning style}” and P120 pointed out “\textit{More controllability is always a good thing, I guess because you are more in charge and feel more able in that sense. I love being able to customize my recommendations}”. 

In the \textit{output control} condition, participants described recommendation quality as improved by sorting, filtering, and feedback options. For example, P144 emphasized, “\textit{Yes absolutely, it meant I wasn't seeing items that were of no use to me, and it helped me with items that are of use to me}”. Similarly, P149 remarked, “\textit{Yes, because if it was less controllable then I wouldn’t be able to determine what type of videos would be shown to me and they would be more likely to be random/less helpful}”. Others highlighted narrowing results as key to quality, as P160 explained, “\textit{It allows me to narrow down results to be as relevant as possible by adjusting the many given settings shown}”. Finally, P155 linked quality to personalization as, “\textit{Yes, it is a more personal learning experience when you can control and change, if feels you can deep dive into the information}”. Our results align with findings from the RS literature concluding that providing control with the output leads to increase in perceived quality of recommendations \cite{smallworlds, chen2012cofeel} and that lower levels of user control with the output negatively influence the perceived quality of recommendations \cite{maxwell2015}.

Overall, our results show that perceived recommendation quality was consistently rated high across all conditions, which indicates that user control at different levels positively influences perceived recommendation quality. 
Moreover, participants especially appreciated the customization options, which gave them confidence that the system would deliver relevant, tailored, and accurate results.

In summary, our findings suggest that even in the absence of explicit control
features, participants perceived the ERS as controllable, transparent, trustworthy, satisfactory, accurate, and useful, primarily because
the recommendations were grounded in their own user profile that they created themselves (i.e., the “not understood”
concepts they had marked). This implicit form of control over the user profile was sufficient to create strong baseline
perceptions of agency and control. Additional control mechanisms, while often desired to add/remove or adjust the DNU concepts or
refine the recommendation results, appear to serve more as enhancements than necessities. Thus, providing control
over the user profile alone can already foster a positive perception of the ERS, while additional control features can offer
opportunities to strengthen, but not fundamentally build these perceptions. The fact that many participants desired
more control over the ERS \textit{input} implies that allowing users to edit their user profile can lead to increased perceived
control. Consequently, we suggest that, empowering users to create and edit their user profile should be an integral
feature in any ERS to foster users’ perceived control, transparency, trust, satisfaction, and perceived recommendation
quality in the ERS.
\subsection{Impact of User Control on Different Goals 
and the Relation between them} \label{impact}
To examine RQ2 ("what are the effects of user control on transparency of, trust in, satisfaction with, and perceived recommendation quality of the ERS, and how do these goals relate to each other?),
we computed Pearson Correlation between user control and the recommendation goals of transparency, trust, satisfaction, and perceived recommendation quality.
We first examined the relationships between all these DVs across all participants (N = 184), irrespective of control condition, to obtain an overall picture of how perceived control, transparency, trust, satisfaction and perceived quality are interrelated. 
Moreover, we investigated the relationships between all these DVs for each control condition. 
For each correlation, 95\% confidence intervals were estimated using Fisher’s z transformation. 
As shown in Figure \ref{heatmapall} and Figure \ref{fig:four}, all the constructs user control, transparency, trust, satisfaction, and perceived quality were positively correlated with statistical significance, though with varying strengths, suggesting that these constructs are intertwined, and improvements in one aspect can positively impact the others. 
To further examine the directional and multivariate relationships among these constructs, we conducted a structured equation modeling (SEM) analysis. Figure \ref{sem} shows the results of the SEM analysis using composite construct scores as observed variables, showing $R^2$ and path loadings. As all of the $R^2$ estimates are larger than 0.10, they are appropriate and informative to examine the significance
of the paths associated with these variables \cite{Pu2011resque}. While the correlation analysis provided evidence of significant associations between the recommendation goals, SEM enables testing directional relationships and examining these constructs simultaneously within a single model. The model was specified to reflect a sequential evaluation process in which perceived control influences transparency, transparency shapes perceived quality, perceived quality strengthens trust, and trust subsequently affects user satisfaction. The model was estimated using maximum likelihood with bootstrapped standard errors (5,000 resamples), and exhibited a strong fit\footnote{Although the $\chi^2$ test is significant ($p = .014$), this statistic is sensitive to sample size and model complexity \cite{bentler1980significance}. While we used the standard cutoff criteria by \citet{byrne1994structural}(e.g., CFI and TLI > .90, RMSEA < .05), RMSEA and TLI are known to be unreliable when df is very small \cite{kenny2015rmsea}. Accordingly, model fit is primarily evaluated using CFI (.992) and SRMR (.024), both indicating good fit.} to the data across several indices
$\chi^2(1) = 6.09$, $p = .014$, 
(CFI = .992, TLI = .916, and SRMR = .024). Although the RMSEA value (RMSEA = .166, 90\% CI [.060, .303]) was above the preferred threshold, this index is known to be unreliable and artificially inflated in models with low degrees of freedom (df = 1) \cite{kenny2015rmsea}.
The model explained a substantial proportion of variance in the constructs ($R^2$ values), including 39.4\% of the variance in transparency, 43.4\% in perceived quality, 57.2\% in trust, and 68.6\% in satisfaction.
\subsubsection{Impact of User Control on Different Goals} \label{impact-control-goals}
Across all participants, perceived user control was strongly correlated with transparency (r = .63, p < .001) and satisfaction (r = .61, p < .001), and showed a moderate association with trust (r = .57, p < .001) and perceived quality (r = .55, p < .001). This indicates that participants who felt more in control also tended to perceive the system as more transparent, trustworthy, and satisfying. 
When examined per control condition, these associations were consistent but varied in strength. 

Across all participants, perceived user control was strongly associated with \textit{transparency} (r = .63, p < .001). 
The SEM results further highlight that perceived control had a strong positive effect on transparency ($\beta = .63$, $p < .001$).
This suggests that participants who felt more in control also perceived the system as clearer and more understandable, as one important aspect that may contribute to increased transparency is the degree of control users have over the system \cite{knijnenburg2012inspectability, he2016interactive, shneiderman2022human}. One reason we deem responsible for the high level of transparency among the majority of participants is that they were able to control the ERS in a way that best fits their needs and preferences. 
Consequently, we argue that, empowering users to take control of the recommendation should be an integral feature in any ERS, if transparency is a desirable property for the ERS. We refer to this type of transparency as transparency through controllability. 
Therefore, our study provides evidence that user control with the ERS, leads to an increased transparency. This confirms findings from previous studies from the recommendation domain which pointed out that user control is very closely tied to  transparency \cite{tintarev2015explaining} and that control significantly affects transparency \cite{Pu2011resque, pu2011user}. Similarly, in their study, \citet{hellmann2022development} also found that transparency with respect to interaction is rated
higher when users report
higher perceived control. This is also in line with studies from interactive recommendation literature which show that user control can also contribute to increased transparency of RSs \cite{tsai2021effects, tsai2017providing, knijnenburg2012explaining}.
The condition-level analysis showed that this strong correlation was also found in the \textit{basic control} (r = .61), \textit{input control} (r = .72), and \textit{process control} (r = .64) control conditions. Only in the \textit{output control} condition the association between user control and transparency was moderate (r = .55). This result is consistent with our qualitative findings, showing that control with the input and process was found to be contributing more to the transparency of the ERS, compared to control with the output. One possible reason could be that transparency (i.e., the extent to which information about a system’s
reasoning is provided and made available to users \cite{guesmi2023justification}) is more related to the \textit{input} or \textit{process}, rather than the \textit{output} of the ERS. Transparency can be
distinguished with respect to input (what data does the system use?) or process (how is an item recommended?) {\cite{hellmann2022development}.    
By contrast, controlling the output (i.e., recommendation results) through filtering, sorting, and rating does not necessarily lead to understanding how the system works and thus does not lead to a higher perception of system's transparency.  
Perceived user control correlated moderately with \textit{trust} in the overall sample (r = .57, p < .001). However, the SEM results indicate that the relationship between control and trust appears to operate indirectly through other goals. In particular, the path [control $\rightarrow$ transparency $\rightarrow$ trust] yields an indirect effect of $0.16$ ($p=.001$), while [control $\rightarrow$ perceived quality $\rightarrow$ trust] has an indirect effect of $0.155$ ($p=.009$). In addition, the sequential path [control $\rightarrow$ transparency $\rightarrow$ perceived quality $\rightarrow$ trust] yields an indirect effect of $0.18$ ($p <.001$). 
Our finding related to the indirect effect of control on trust closely aligns with prior work by \citet{ooge23steering}, who investigated the effect of control mechanisms on trust in an ERS. Their results showed that providing users with control over recommendations, primarily at the input level, did not significantly increase trust on its own, but rather encouraged users to reflect on their mastery level and the underlying recommendation algorithm. Similarly, in our model, the direct effect of control on trust is limited. Our study showed that its influence is primarily indirect through transparency and perceived quality. Concretely, we found that perceived control significantly increases transparency and perceived quality, both of which subsequently contribute to users’ trust in the system. When users can observe how their interactions influence the recommendations, they may perceive the system as more transparent and the recommendations as more relevant, which in turn fosters trust. This suggests that control contributes to trust by helping users better understand how the ERS works and by enabling them to evaluate the relevance and usefulness of the recommendations. Consistent with our qualitative findings (see Section \ref{control_trust}), these results provide evidence that in the interactive RS domain, similar to
findings from the explainable RS domain \cite{siepmann2023trust,yang2020visual,miller2022we,liao2022designing}, trust is a multidimensional concept which is influenced by multiple factors. Condition-specific correlation analyses revealed that the correlation between user control and trust varied across levels of control. In the \textit{basic control} and \textit{output control} condition, user control was strongly correlated to trust (r = .61 and r = .67, respectively). 
Whereas, under \textit{input} (r = .50) and \textit{process} (r = .54) control, the association is moderate.
This positive correlation between user control and trust confirms findings in other studies in the RS domain highlighting that providing more control over the RS is an important strategy to secure trust \cite{harambam,jannach2017user,jannach2019explanations} and that interactive features that allow users to control the RS increase their trust in the system \cite{schaffer2015hypothetical, bruns2015should, loepp2014choice}.
Our results further indicate that the relationship between user control and trust was weaker in the Input and Process control conditions than the \textit{basic control} and \textit{output} conditions. This confirms that trust depends primarily on the RS output and its ability to formulate good
recommendations (i.e., perceived usefulness) \cite{Pu2011resque} and
the accuracy of the recommendation algorithm \cite{tintarev2011}.
Another possible reason can be that, as found in our qualitative results, the control mechanisms provided in the Process condition can be overwhelming for users which might negatively impact their trust in the ERS. 
This indicates that simple interactions that are familiar to users like filtering, sorting, and rating as provided in the Output condition can be enough to foster trust. 

Our study further shows that user control correlated strongly with \textit{satisfaction} overall (r = .61, p < .001). Moreover, the SEM analysis showed a smaller but significant direct effect of user control on satisfaction ($\beta = .13$, $p = .002$). In addition, user control exerted several significant indirect effects, including [control → perceived quality → satisfaction] ($\beta = .06$, $p = .027$), [control → transparency → perceived quality → satisfaction] ($\beta = .07$, $p = .009$), [control → perceived quality → trust → satisfaction] ($\beta = .07$, $p = .017$), and [control → transparency → trust → satisfaction] ($\beta = .08$, $p = .004$). Finally, the full sequential pathway [control → transparency → perceived quality → trust → satisfaction] yielded the strongest indirect effect ($\beta = .08$, $p < .001$). These findings indicate that the effect of user control on satisfaction operates not only directly but also through a chain of mediators involving transparency, perceived quality, and trust. 
Our results showing a positive relationship between user control and satisfaction confirm the findings by \citet{Pu2011resque}, who noted that user control weighs heavily on the overall satisfaction with the RS. Similarly, \citet{knijnenburg2012inspectability} pointed out that, by providing control over various stages of the RS, users can tailor their experience more closely to their personal preferences, thereby enhancing satisfaction. 
Furthermore, several studies from the interactive recommendation domain evidenced the positive effects of user control on satisfaction \cite{tsai2017providing, linkedvis2013, smallworlds, talkexplorer, setfusion2014, wong2011diversity, 2012tasteweights} and user experience \cite{Donovan, 2012tasteweights, setfusion2014, schaffer2015hypothetical,parra2015user}. 
In the condition-wise correlation analysis, the strongest correlation was found in the \textit{basic control} condition (r = .74). Also the \textit{output control} showed a strong association (r = .68). Whereas, under \textit{input} (r = .50) and \textit{process }(r = .53) control, the association was moderate. 
One reason that we deem responsible for this result is that participants were
inclined to be satisfied with the system when 
it provides good recommendations that match their interests. This suggests that, to achieve user satisfaction, further interaction with the input and process of the ERS is not needed if the system performs as expected.    

Our study further shows that user control correlated moderately with \textit{perceived quality} in the overall sample (r = .55, p < .001). Moreover, the SEM analysis revealed both significant positive direct and indirect effects of control on perceived quality. The direct effect was ($\beta = .26$, $p = .004$), suggesting that having control over the system may help users better evaluate the accuracy and usefulness of the recommended items. 
This finding aligns with prior work suggesting that providing users with control over RSs can enhance their perception of the recommendation quality \cite{knijnenburg2012inspectability, jin2016go, tsai2021effects}, accuracy \cite{schaffer2015hypothetical, Donovan, maxwell2015, smallworlds, 2012tasteweights}, and usefulness of recommendations \cite{Pu2011resque, Kangasrasio2015, zhao2010, chen2012cofeel, tintarev2015inspection, talkexplorer}.
The indirect path [control $\rightarrow$ transparency $\rightarrow$ perceived quality] yields an effect of $0.29$ ($p<.001$). This indicates that perceived quality is not only influenced directly by perceived control but also indirectly through transparency. This aligns with findings by \citet{knijnenburg2012inspectability} noting that user control has a positive effect on the understandability 
of the RS, which in turn increases the perceived quality of the recommendations. Condition-wise correlation analysis showed that the strongest association between control and perceived quality was observed in the Output control condition (r = .67), followed by the Process (r = .60), Input (r = .54), and Basic control (r = .49) conditions. This pattern suggests that when users can directly interact with the recommendation results or influence the recommendation process, their perception of recommendation quality becomes higher. 
One possible explanation is that interactive control that allows users to adjust the recommended items (i.e., output) or the system behavior (i.e., process) will lead to re-ranking the items. This allows the most relevant items to appear at the top of the recommendation list, which can strengthen the user's perception of recommendation quality. 

In summary, our findings clearly indicate that perceived user control plays an important role in shaping users’ perceptions of the ERS. In particular, user control is a key determinant of transparency in IntRSs. 
Moreover, we found that not all levels of control have the same effects on recommendation goals.
\textit{Input} and \textit{process controls} appear to strengthen users’ sense of transparency as they lead to a better understanding of how the system works. On the other hand, \textit{output control} primarily supports trust, satisfaction, and perceived quality, indicating that providing simple control options that users are familiar with and good recommendations that match users' expectations are more important factors to improve trust, satisfaction, and perceived quality. 
Taken together, these findings highlight the complementary roles of different levels of control.
From a design perspective, we suggest that IntRSs should aim for a balanced provision of control features, combining input-level and process-level controls to foster transparency with output-level controls to enhance trust, satisfaction, and perceived quality.
\begin{figure}[!ht]
\includegraphics[width=0.6\linewidth]{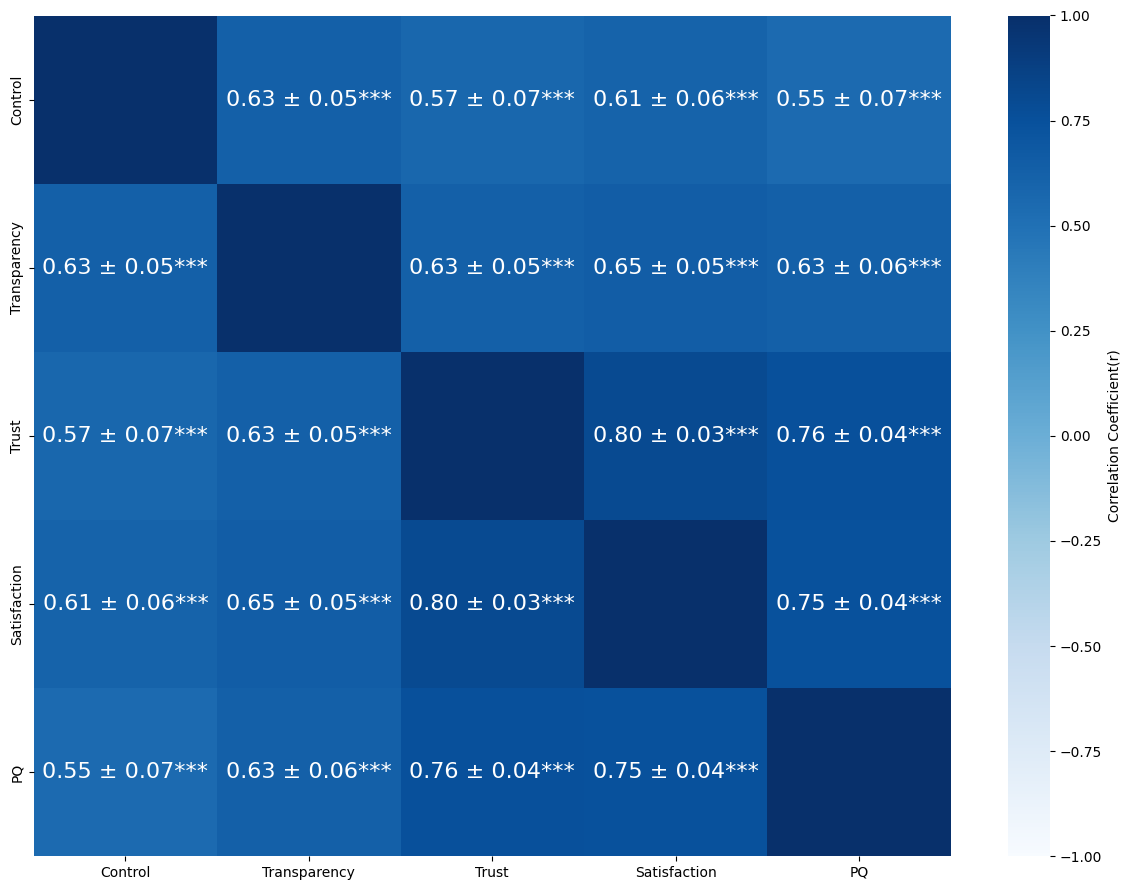}
\caption{Pearson Correlation between different goals, with 95\% confidence interval, where statistically
significant correlations are marked with an asterisk (Significance levels: *** p < .001, ** p < .01, * p < .05.).}
\label{heatmapall}
\end{figure}
\begin{figure}[!ht]
\includegraphics[width=0.8\linewidth]{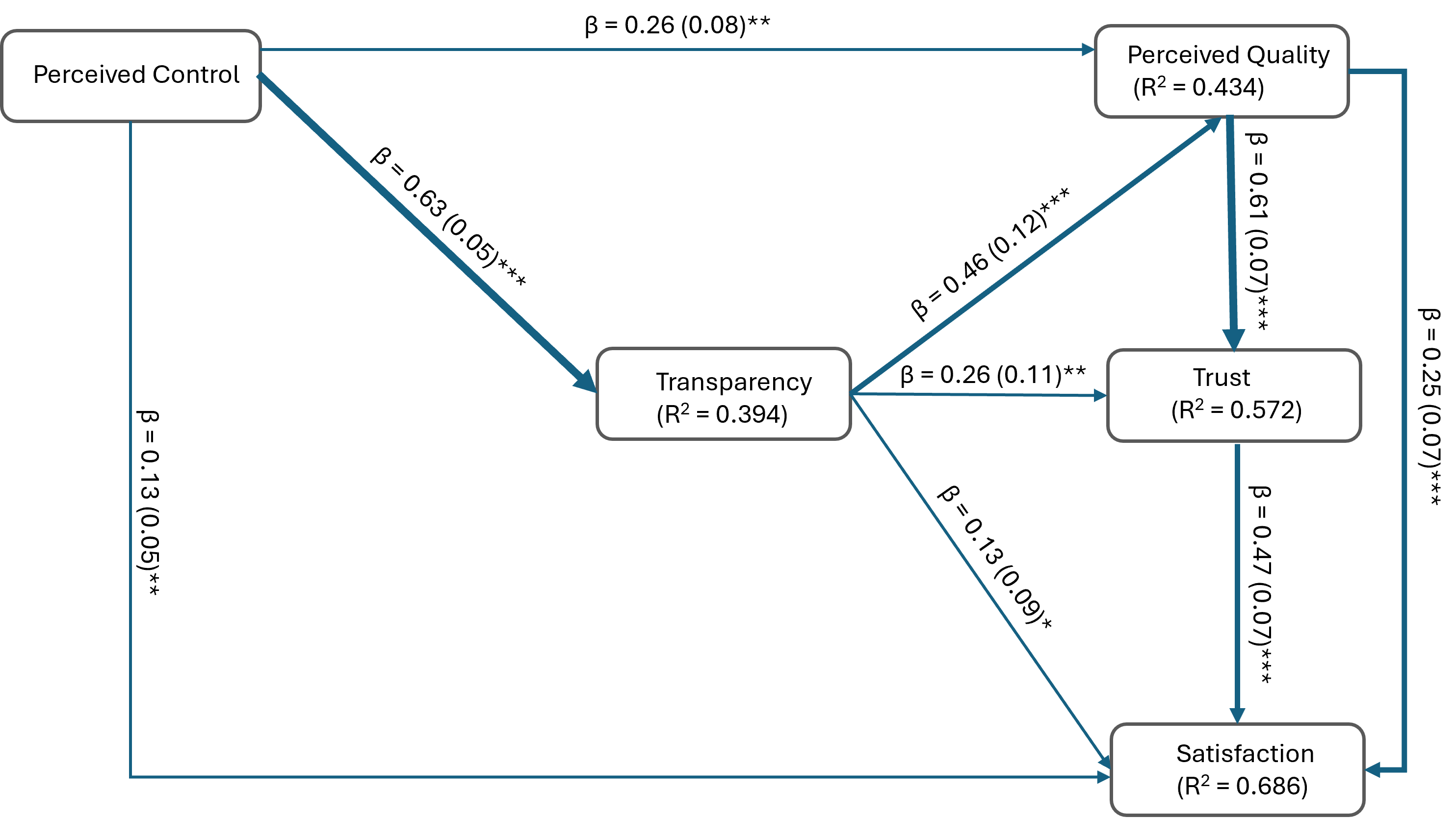}
\caption{Structural path model with standardized coefficients ($\beta$) and standard errors in parentheses. Arrow thickness reflects effect magnitude. Significance levels: *** p < .001, ** p < .01, * p < .05.}
\label{sem}
\end{figure}
\begin{figure}[t]
  \centering
  \begin{subfigure}[b]{0.48\textwidth}
    \centering
    \includegraphics[width=\linewidth]{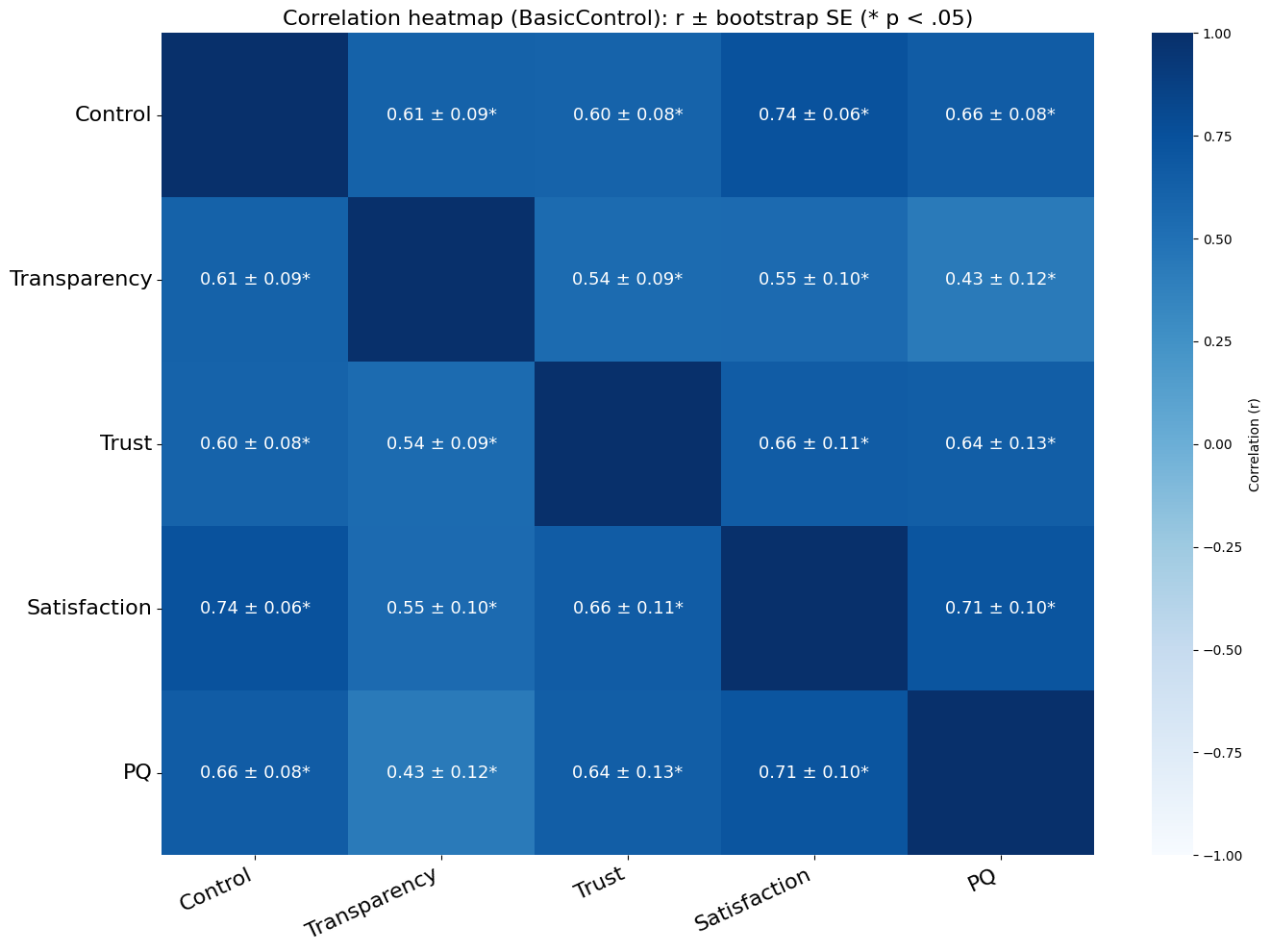}
    \caption{Basic Control}
    \label{fig:four-a}
  \end{subfigure}\hfill
  \begin{subfigure}[b]{0.48\textwidth}
    \centering
    \includegraphics[width=\linewidth]{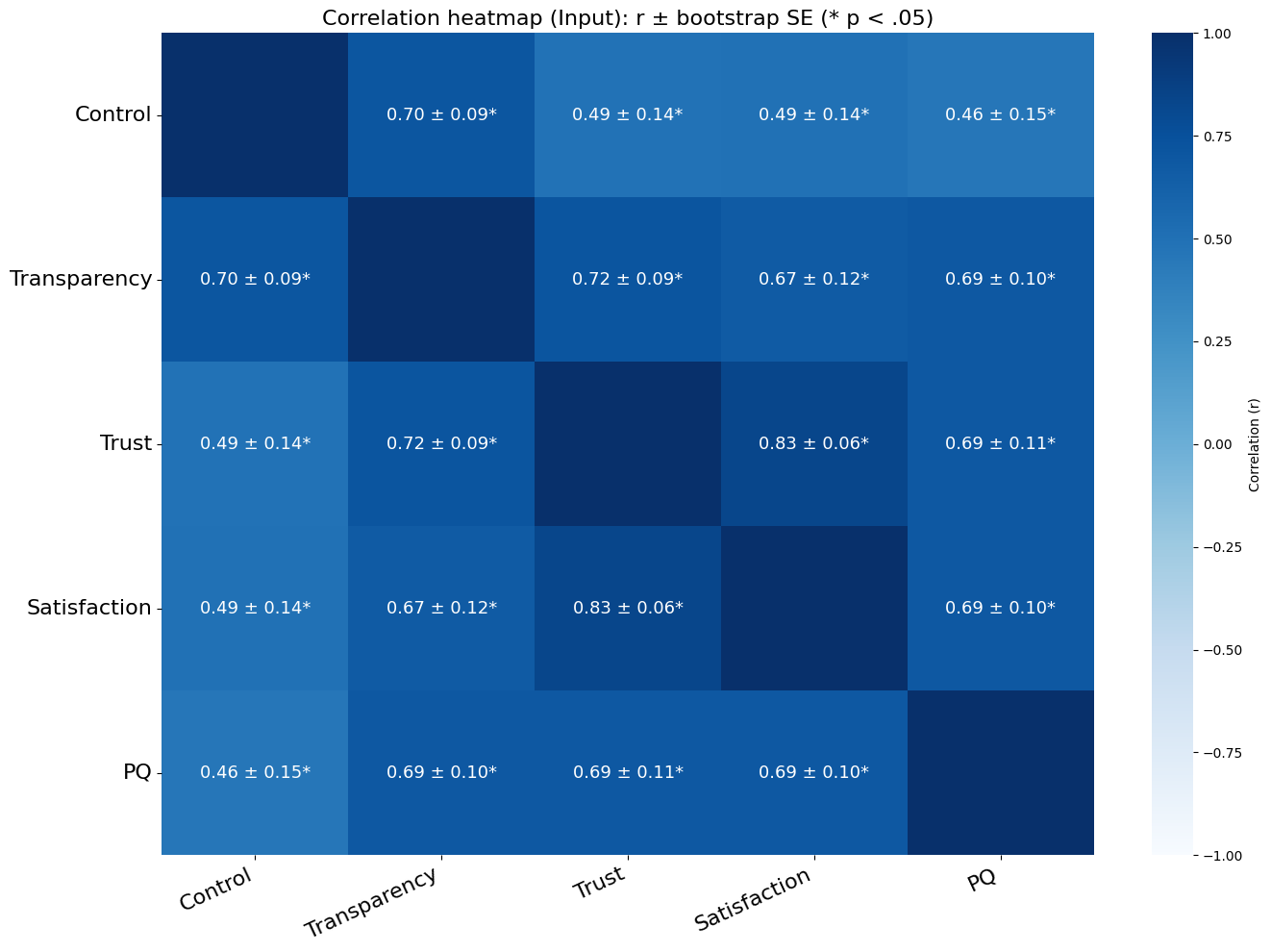}
    \caption{Input Control}
    \label{fig:four-b}
  \end{subfigure}
  \vspace{0.6em} 
  \begin{subfigure}[b]{0.48\textwidth}
    \centering
    \includegraphics[width=\linewidth]{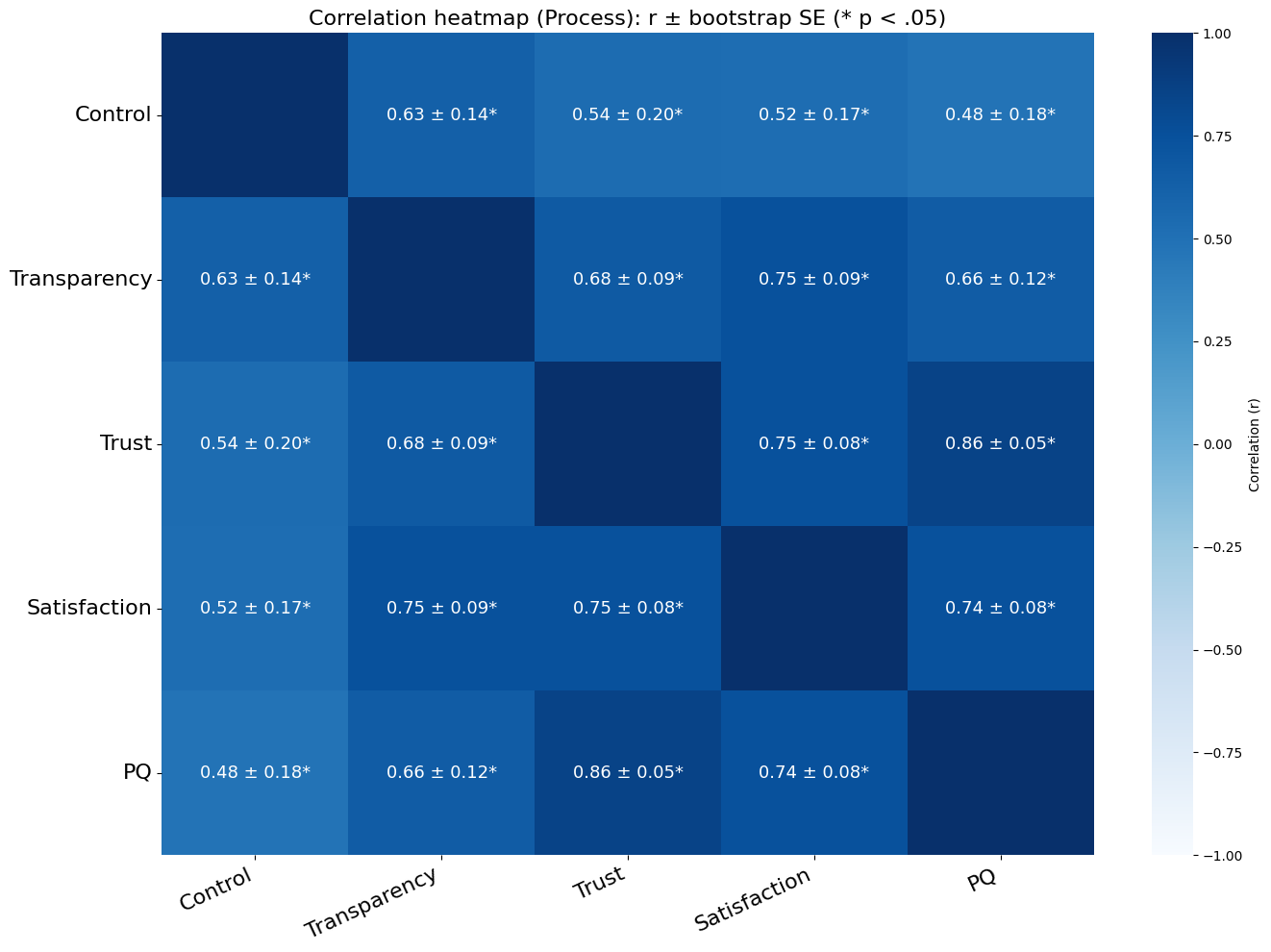}
    \caption{Process Control}
    \label{fig:four-c}
  \end{subfigure}\hfill
  \begin{subfigure}[b]{0.48\textwidth}
    \centering
    \includegraphics[width=\linewidth]{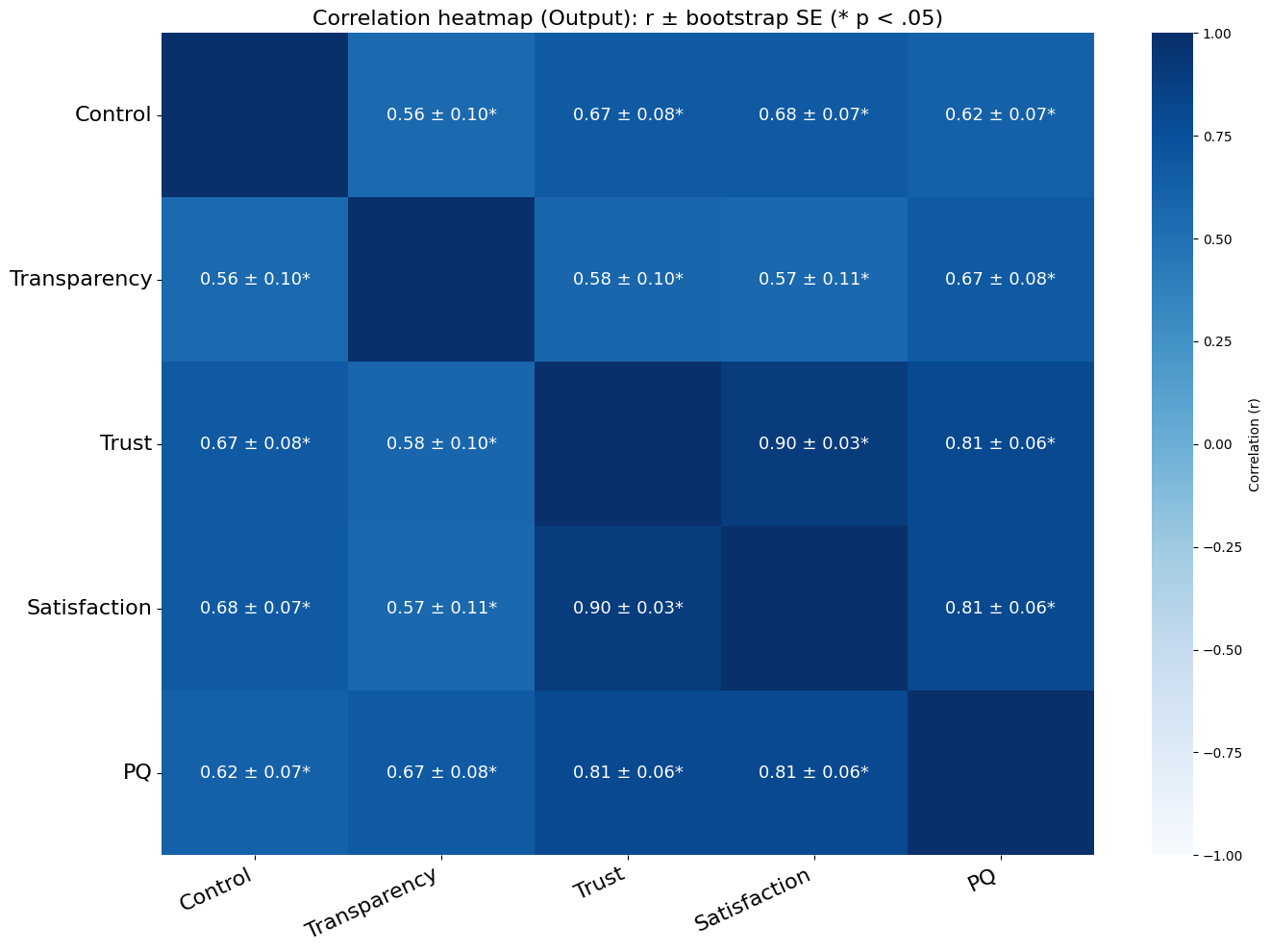}
    \caption{Output Control}
    \label{fig:four-d}
  \end{subfigure}
  \caption{Condition-wise Correlation analysis between different goals, with 95\% confidence interval, where statistically
significant correlations are marked with an asterisk (*)}
  \label{fig:four}
\end{figure}
\subsubsection{Relationships Between Recommendation Goals} \label{interaction-goals}
To investigate how the recommendation goals of transparency, trust, satisfaction, and perceived quality relate to each other (RQ2), we first conducted Pearson Correlation analyses, followed by structural equation modeling (SEM) to examine the directional and multivariate relationships among these constructs. The results from the correlation analysis reveal that all the goals appear to move together, as shown in Figure \ref{heatmapall} and Figure \ref{fig:four}.
Figure \ref{sem} shows the results of the structural model analysis as well. 

We found in the overall analysis that satisfaction strongly correlates with trust (r = .80, p < .001), perceived quality (r = .75, p < .001), and transparency (r = .65, p < .001) with a statistical significance. SEM further revealed that satisfaction was influenced by multiple factors. Trust had a significant positive effect on satisfaction ($\beta = .47$, $p < .001$). Perceived quality also directly contributed to satisfaction ($\beta = .307$, $p < .001$). Additionally, transparency had a direct influence on satisfaction ($\beta = .13$, $p = .049$).
Beyond these direct effects, several significant indirect effects were observed. Transparency influenced satisfaction indirectly through trust [transparency → trust → satisfaction] ($\beta = .12$, $p = .003$) and through perceived quality [transparency → perceived quality → satisfaction] ($\beta = .12$, $p = .005$). Moreover, a sequential pathway [transparency → perceived quality → trust → satisfaction] yielded an indirect effect of $\beta = .13$ ($p = .001$). In addition, perceived quality had a significant indirect effect on satisfaction through trust [perceived quality → trust → satisfaction] ($\beta = .28$, $p < .001$). These findings highlight that satisfaction is not only directly shaped by transparency, trust, and perceived quality, but also indirectly influenced through interconnected relationships among these constructs.
Moreover, in condition wise analysis, we observe the same pattern for strong correlation between satisfaction and trust in all conditions. 
The strongest satisfaction-trust association was observed in the \textit{output control} condition (r = .90). One possible reason could be that when users are able to rely on the ERS to understand a difficult concept, they tend to be satisfied with the recommendation results. 
The satisfaction-transparency association was strongest in the \textit{process control} condition (r = .76), followed by the \textit{input control} condition (r = .69). In the \textit{output} (r = .57) and \textit{basic control} (r = .55) conditions, the association was moderate. 
This pattern suggests that transparency is more related to input and process controls and that users tend to be satisfied with an ERS if its inner working is transparent to them. Our results are in line with prior research suggesting that overall user satisfaction with RS is related to transparency and trust. For instance, in the explainable RS domain, \citet{balog2020measuring} and \citet{guesmi2023interactive, guesmi2023justification} found that satisfaction is positively correlated with transparency and trust. In the same context, \citet{gedikli2014should} reported results from experiments with different explanations clearly showing that transparency has a significant positive effect on user satisfaction. Moreover, a lower-transparency RS is known to negatively affect user satisfaction with the RS \cite{tintarev2015explaining}.
Furthermore, the satisfaction-perceived quality association was strongest in the \textit{output control} condition (r = .81), followed by the \textit{process} (r = .74), \textit{basic control} (r = .71) and the \textit{input control} (r = .69) conditions. 
These findings suggest that perceived recommendation quality plays an important role in shaping user satisfaction across all conditions.
This observation is consistent with prior research indicating that the perceived relevance and usefulness of recommended items are key determinants of user satisfaction in RSs \cite{knijnenburg2012explaining, tsai2021effects,Pu2011resque}}. 
Overall, our results provide evidence that, similar to  findings from the explainable RS domain, transparency, trust, and perceived quality have significant positive effects on user satisfaction in the interactive ERS domain as well. 
Moreover, we extend this evidence to different levels of control in an ERS, indicating that, in particular, trust and perceived quality are key determinants of user satisfaction in IntRSs at all control levels and transparency achieved through controllability can meaningfully contribute to satisfaction, especially when participants engage with the system at the input or process level. 

Our study further shows that the correlation between transparency and trust was strong in the overall sample (r = .63), indicating that transparency and trust are linked. SEM further confirmed that transparency had a smaller but significant direct effect on trust ($\beta = .26$, $p = .005$). These results confirm findings by \citet{hellmann2022development} highlighting that, if users can understand the
internal mechanisms, they trust the RS more. The authors included the effects of
transparency on trust in their development of a questionnaire to measure the perceived transparency in an RS.
They found a positive link between transparency and trust-related measures, and that providing transparency could enhance
users’ trust in the system. \citet{tintarev2015explaining} also 
pointed out that an increase in transparency should
lead to an increase in trust. 
This is also consistent with results from \citet{kizilcec2016much} who found that if  transparency is provided, it fosters trust. 
Similarly, \citet{nunes2017systematic} identified transparency as one of the key factors for users
to develop trust. In addition to this direct relationship, transparency also influenced trust indirectly through perceived quality. Specifically, the path [transparency → perceived quality → trust] yielded a significant indirect effect ($\beta = .28$, $p < .001$), indicating that transparency enhances trust not only directly but also by improving users’ perception of recommendation quality. 
Condition-wise correlation analysis further reveals that the correlation between transparency and trust was less consistent across conditions compared to other associations. It was strongest in the \textit{input control} (r = .72) and \textit{process control} (r = .68) conditions, suggesting that when participants could control input or influence the recommendation process, they were able to understand how the system works which increased their trust. Moreover, we found that transparency and trust stand out as less correlated with each other in the \textit{basic control} (r = .55) and \textit{output control} (r = .58) conditions. This finding indicates that transparency does not automatically translate into higher trust, particularly when users' influence is limited to the ERS output.
One reason that can explain the relatively low correlation between transparency and trust in the Output condition is that while transparency can be achieved through controllability, interacting with the output of the RS cannot always assure that users understand the underlying rationale of the RS, especially when the recommendation mechanism is too complicated to non-expert users \cite{tsai2021effects}, which can negatively impact the perceived trust of the RS.  
Some considerable transparency could be achieved through explanation which is also recognized as an important factor that fosters user trust in RS, as it can improve users' understanding of how the system works \cite{tintarev2015explaining}.
We hypothesize that a different result may be observed if explanations were provided in the ERS together with interactions. It is therefore important to explore in the future work the effects of different types of transparency (i.e., transparency through controllability vs. transparency through explanation) on users’ trust in RSs \cite{siepmann2023trust,chatti2023visualization}.

Our correlation analysis further reveals a strong positive association between perceived quality with both trust (r = .76) and transparency (r = .63) in the overall sample. SEM results further confirm these findings where perceived quality significantly influences trust in the ERS ($\beta = .605$, $p < .001$), indicating that users who perceive the recommendations as more relevant and useful are substantially more likely to trust the system. 
This relationship can be explained through the user's trusting belief in the ERS’ competence. 
According to \citet{mcknight2002developing}, competence belief refers to users’ perception that a system has the ability, skills, and expertise to perform
effectively. 
In our study, when users perceive the recommendations as relevant and well-matched to their interests, they are able to form trusting
beliefs concerning whether or not the ERS has the expertise to provide accurate and useful recommendations (i.e., is competent). Our results suggest that higher perceived quality can strengthen users’ trusting belief in the ERS' competence.
This aligns with prior empirical work showing that trust development is influenced by recommendation quality. In their study, \citet{kunkel2019let} found that recommendation quality has the strongest influence on perceived competence of the RS. Similarly, a recent study by \citet{saxborn2024trust} also 
highlights the crucial impact of perceived
recommendation quality in shaping trusting beliefs. The authors found that users tend to associate high recommendation quality with greater trust in the RS, while poor recommendations can lead to distrust with the system.
The condition-wise correlation analysis further revealed that the strongest correlation between perceived quality and trust was observed in the \textit{process control} condition (r = .86), followed by the \textit{output} (r = .81), \textit{input} (r = .69), and \textit{basic control} (r = .64) conditions. This indicates that, across all conditions, perceived recommendation quality is a key determinant of user trust in the ERS. 

Referring to the relationship between perceived quality and transparency, SEM showed that transparency acted as a vital construct in predicting perceived quality ($\beta = .46, p < .001$) indicating that users who perceive the system as more understandable also tend to evaluate the recommendations as more accurate and useful. 
A closer look at the items used to measure transparency and perceived quality in our study helps to explain the relationship between them. The transparency construct evaluated users' understanding of how the system works and how the system chooses relevant videos for them (e.g., I can tell how well the recommendations match my preferences). Whereas, the perceived quality construct included items related to the relevance of the recommendations based on users' interests (e.g., the recommended videos fitted my preferences). Because transparency enables users to understand how recommendations are generated, they are better able to assess whether or not the recommendations align with their preferences, which enhances their perceived recommendation quality. 
This explanation is also consistent with prior findings suggesting that users evaluate the accuracy of recommendations based on whether they can see a correspondence between their expressed preferences and the recommendations generated by the system \cite{simonson2005determinants}. 
This result is also inline with findings from \citet{tsai2021effects} that perceived transparency leads to an increased perception of recommendation quality in IntRSs. 
However, the strength of this association varies across conditions. The weakest correlation between transparency and perceived quality was observed in the \textit{basic control} condition (r = .43), while stronger correlations were found in the \textit{input} (r = .69), \textit{output} (r = .67) and \textit{process} (r = .66) control conditions. This pattern suggests that basic control of the ERS (i.e., knowing what data does the system use) is insufficient for users to effectively assess the alignment between recommendations and their preferences. In contrast, transparency at the input level (i.e., adjusting what data does the system use), process level (i.e., how and why is an item recommended?), and output level (i.e., why and how well does an item fit one’s preferences?) appears to better support users in perceiving this alignment. This, in turn, facilitates a more informed evaluation of recommendation quality.

In conclusion, our results reveal a strong interconnection between the recommendation goals. 
In particular, satisfaction is strongly correlated with trust, perceived quality, and transparency. Notably, users tend to be satisfied with an ERS if its inner working is transparent to them through input and process controls.
Transparency is positively associated with perceived quality and trust, suggesting that helping users understand how recommendations are generated can positively impact users' perception of recommendation accuracy and usefulness and increase their confidence in the system. In particular, transparency is a strong driver of perceived quality in IntRSs.
Additionally, while transparency through controllability of the ERS output may support trust, it is not sufficient on its own, as trust also depends on other factors, such as providing explanations.
Furthermore, our results show that in the IntRS domain, trust is a multi-faceted concept which is influenced by transparency and perceived quality, and that recommendation quality has the strongest influence on trust in the ERS. Overall, these findings highlight that transparency, perceived quality, trust, and satisfaction form a closely related set of goals in the ERS, where improvements in one dimension can positively influence the others.
\section{Limitations} \label{limitations}
As a first investigation of the effects of different levels of user control in an interactive ERS, this study is not without limitations. The study was conducted in the MOOC platform CourseMapper, and within a short-term, single-session setting. While this provided a controlled environment for systematic testing, it remains to be seen whether the findings generalize to other educational environments beyond MOOCs, such as learning management systems (LMSs) or intelligent tutoring systems (ITSs), or to longer-term use where perceptions of control, trust, transparency, satisfaction, and perceived quality may evolve over time. Furthermore, the observed effects were found based on the specific user control mechanisms implemented in the study system. Alternative implementations (e.g., richer filtering options or interactive explanations) may yield different user experiences and outcomes. Moreover, the current study relies on self-reported user perceptions and does not include their interaction behaviors which could help contextualize participants’ reported perceptions. Finally, while our measures captured participants’ perceptions, they did not include direct assessments of learning performance, making it unclear to what extent perceived control, transparency, trust, and satisfaction translate into improved educational outcomes.
\section{Conclusion and Future Work} \label{concl}
This paper contributes to a deeper understanding of providing different levels of user control in educational recommender systems (ERSs), an area that remains underexplored in the existing literature. To this end, we systematically designed and evaluated user control at different levels of the ERS, namely with the input (i.e., user profile), process (i.e., recommendation algorithm), and output (i.e., recommendations). Following a mixed-methods evaluation approach, we conducted an online user study (N=184) to explore the impact of providing control at various levels in an ERS on users’ perceived control, transparency, trust, satisfaction, and perceived quality. Moreover, we investigated the relations among these recommendation goals. 

Our findings suggest that providing control over the user profile is sufficient to establish positive perceptions of an ERS, while additional control options serve mainly to reinforce these impressions.
Importantly, perceived control emerges as the only recommendation goal significantly impacted by the level of control, with input control exerting the strongest influence.
Our results further highlight that different levels of control affect the perceptions of transparency, trust, satisfaction, and perceived quality differently and that these recommendation goals are strongly interrelated. 
Overall, this work contributes empirical evidence that user control has a positive influence on transparency, trust, satisfaction, and recommendation quality, albeit to different degrees. The
findings presented here also provide starting points for
research into further elucidating the effects of different levels user control in ERSs.  
Future research should further explore hybrid approaches that combine controllability with explanations, examining how transparency through controllability and transparency through explanation can complement one another in educational contexts. Another possible future improvement is adaptive control mechanisms, where the system adjusts the level of control based on user's personal characteristics. In addition, future research should investigate how users’ perceived experience (e.g., transparency, trust, and perceived quality) translates into actual interaction behavior and learning-related outcomes.

\bibliographystyle{ACM-Reference-Format}
\bibliography{sample-base}
\end{document}